\documentclass[10pt, journal]{IEEEtran}

\usepackage{graphicx}
\usepackage{enumitem}
\usepackage{amsmath}
\usepackage{breqn}
\usepackage{pbox}
\usepackage{multirow}
\usepackage{color}
\usepackage{subfig}
\usepackage{url}
\usepackage{hyperref}
\usepackage{verbatim}
\usepackage{tabularx}
\definecolor{bluebell}{rgb}{0.64, 0.64, 0.82}
\usepackage{listings,xcolor}
\ifCLASSOPTIONcompsoc
  
  \usepackage[nocompress]{cite}
\else

  \usepackage{cite}
\fi 

\ifCLASSINFOpdf
\else
\fi

\newcommand\inlineeqno{\stepcounter{equation}\ (\theequation)}

\begin{document}
\setlength{\parindent}{0pt}

\title{A Policy based Security Architecture for Software Defined Networks}
\author{Vijay Varadharajan,~\IEEEmembership{Senior Member,~IEEE,} 
 Kallol Karmakar,~\IEEEmembership{Student Member,~IEEE,} and
        Uday Tupakula,~\IEEEmembership{Member,~IEEE}  and
       Michael Hitchens
\thanks{Vijay Varadharajan is with the Faculty of Engineering, The University of Newcastle, Australia. E-mail: vijay.varadharajan@newcastle.edu.au}
\thanks{Kallol Karmakar and Uday Tupakula are with the School of Electrical Engineering and Computing, The University of Newcastle, Australia. E-mail: kallolKrishna.Karmakar@newcastle.edu.au and uday.tupakula@newcastle.edu.au}
\thanks{Michael Hitchens is with the Dept of Computing, Faculty of Science and Engineering, Macquarie University, Australia. E-mail: michael.hitchens@mq.edu.au}

\vspace{-8mm}

}
\IEEEtitleabstractindextext{
\begin{abstract}
As networks expand in size and complexity, they pose greater administrative and management challenges. Software Defined Networks (SDN) offer a promising approach to meeting some of these challenges. In this paper, we propose a policy driven security architecture for securing end to end services across multiple SDN domains. We develop a language based approach to design security policies that are relevant for securing SDN services and communications. We describe the policy language and its use in specifying security policies to control the flow of information in a multi-domain SDN. We demonstrate the specification of fine grained security policies based on a variety of attributes such as parameters associated with users and devices/switches, context information such as location and routing information, and services accessed in SDN as well as security attributes associated with the switches and Controllers in different domains. An important feature of our architecture is its ability to specify path and flow based security policies, which are significant for securing end to end services in SDNs. We describe the design and the implementation of our proposed policy based security architecture and demonstrate its use in scenarios involving both intra and inter-domain communications with multiple SDN Controllers. We analyse the performance characteristics of our architecture as well as discuss how our architecture is able to counteract various security attacks. The dynamic security policy based approach and the distribution of corresponding security capabilities intelligently as a service layer that enable flow based security enforcement and protection of multitude of network devices against attacks are important contributions of this paper.
\end{abstract}
}
\maketitle
\IEEEdisplaynontitleabstractindextext
\IEEEpeerreviewmaketitle

\vspace{-6mm}

\section{Introduction}\label{sec:Introduction}
\IEEEPARstart{A}{S} networks expand in size and complexity, they pose greater administrative and management challenges. Increasingly, current networks are highly heterogeneous with many different devices, from small sensors and appliances to network devices such as routers to many different clients and servers and peripherals. Furthermore, these devices use different network technologies such as fixed, wireless and mobile networks. In such a complex heterogeneous environment, management of network devices (such as switches and routers), the mobility of users and devices, the dynamic variation in networks (due to failure of devices and network links), as well as the dramatic increase in security attacks are posing serious challenges.  Software Defined Networks (SDN) \cite{onf} offer a promising approach to meeting some of these challenges. 
SDN is rapidly emerging as a disruptive technology, poised to change communication networks much the same way cloud computing is changing the \lq\lq compute\rq\rq ~world. It is altering the texture of modern networking, moving away from the current control protocols dominant in the TCP/IP Internet stack, towards something more flexible and programmable. It is potentially changing the way networking will be conducted in the future, by enabling devices that are open and controllable by external software, unlike today\rq s proprietary network equipment that has protocols embedded into them by the vendors. 
SDN opens up new avenues of research to realize network capabilities that were impossible or extremely cumbersome before, thereby helping to make future networks more manageable and practicable. The separation of the control plane from the data plane in SDN results in the network switches becoming simpler forwarding devices with the more sophisticated control logic implemented in software in a logically centralized Controller. This decoupling in SDN enables the design of new and innovative network functions and protocols. First, it is simpler and less error-prone to modify network policies through software, than via low-level device configurations. Second, a control program can automatically react to spurious changes in the network state and thus maintain up to date high-level policies. Third, the centralization of the control logic in a Controller with network domain wide knowledge can help to simplify the development of sophisticated network functions. Although SDN offers such advantages to deal with complexities in current networks, a critical issue in SDN at present is that of security; the current state of the art in SDN security is not mature \cite{kreutz13}. Securing networks is becoming more challenging to businesses, especially with bring your own devices (BYOD), increased cloud adoption and the Internet of Things (IoT).

\noindent The main causes of security concerns probably lie in SDN\rq s main benefits, namely programmability of networks and the centralization of control logic. These capabilities introduce new security threats and attack surfaces, which do not exist in traditional networks. Ironically, the closed (proprietary) nature of network switches together with the heterogeneity of vendor software and integrated control functions previously offered “natural” layers of defense in traditional networks. That is, an attack against a network device from a specific vendor will often not work against another network device from another vendor. In some sense, the diversity of network devices and protocols provides a certain level of security, just like secrecy associated with proprietary systems. On the other hand, SDN with its standardized interfaces and protocols (e.g. OpenFlow \cite{onf} protocol between the Controller and the switches) can provide a focused target for the attackers; furthermore, any potential logical security faults in compliant implementations of the SDN protocols and software can increase the security risks. e.g. an attack similar to Stuxnet \cite{langner2011}, which targeted the operation of many devices in specific networked infrastructures by automatically modifying their control programs and configurations, could have dramatic consequences in a highly programmable SDN network. In adddition, the centralization of functionality in the SDN Controller presents another point of security weakness.

\noindent In a SDN environment, the control plane specifies the policies from which the flow rules are derived and enforced in the data plane at the SDN switches and network devices. In this paper, we consider the design of security policies for the establishment of secure end to end communication paths in a distributed SDN environment. In large networks with multiple autonomous system domains crossing organizational boundaries, paths must be selected according to policy related parameters involving security attributes in addition to the traditional parameters of connectivity, congestion and costs. The ability to distribute security capabilities intelligently as a service layer and to have a dynamic security policy based approach to securing a multitude of devices against threats are important contributions of this paper. For instance, we can have policies that enforce the requirement that certain traffic must pass through certain switches which are more secure than others. In this case, a specific set of switches are obligatorily traversed for this traffic. Another example is that to counteract particular threats, certain switches may be forbidden to be in the paths of certain defined traffic. Our policy based architecture combined with the Controller's network domain wide visibility offers a powerful approach to enforcing security mechanisms to counteract security attacks in SDN. This capability is achieved in both the intra and inter-domain contexts with multiple domains managed by different SDN Controllers.

\noindent The main contributions of this paper are as follows: We present the design of a security architecture with specific access policy constraints on communications between end users/devices in a distributed SDN environment with multiple domains. Our method is based on the use of a security policy language based approach enabling secure flow of packets and secure management of paths in a distributed SDN. The security policies are specified at a fine granular level, which when enforced help to detect different types of attack flows, counteracting various threats in the SDN. In our architecture, the security policies use the context associated with the flows, such as location, routing information, services accessed as well as security labels associated with the switches and Controllers, to securely manage the flows. For instance, suppose due to a DDoS attack, traffic from an end host is not able to get through the network. Our SDN security architecture is able to detect the DDoS attack efficiently and establish an alternative path for the traffic from the end host to reach the required destination. Moreover, we can enforce path and flow based policies such as certain communications should go through a path with certain security attributes. Such path based policies are critical when securing data from sensitive applications but are also useful for applications with different quality of service requirements. For instance, video traffic requiring certain bandwidth needs to take a path where the switches and channels have the necessary capabilities (compared to just supporting audio traffic). A novel feature of the proposed security architecture involves the use of the visibility of the network domain and connectivity to manage flow and path based security policies to achieve secure communications and efficient provision of SDN services across multiple domains. Combined with this feature, our architecture has the potential to dynamically update security policies based on the distributed network state and detection of attacks, which is particularly important to counteract emerging security threats. We believe such architectural features are vital for achieving resilient network systems, especially in securing critical infrastructures.

\noindent The paper is organized as follows: After first outlining a brief introduction to SDN, in Section \ref{sec:SATM}, we describe the threat model for our proposed security architecture. We consider the various security threats in a SDN environment such as threats to the Controller, the network switches as well as the communications between the switches and the Controller, and the communications between the Controllers in multiple domains. We then focus on the security threats that our security architecture is designed to address and outline the benefits of our approach. Section \ref{sec:pbsa} describes our proposed policy based security architecture for a distributed SDN. First we give an overview of the proposed architecture. Then, we describe the specification of security policies, which is a key component of the proposed policy based architecture. 
Then we provide a brief walk-through of the security architecture using an inter-domain scenario with multiple domains and SDN Controllers. Section \ref{sec:SAI} describes the implementation of the security architecture and its various components. Section \ref{sec:R} presents the implementation and the performance results. We discuss security policies in both intra and inter domain contexts and illustrate them with example scenarios. Security analysis of some of the implemented scenarios is given in Section \ref{sec:SA}. In Section \ref{sec:RW}, we review the relevant related works and compare our architecture with existing solutions. Finally Section \ref{sec:C} concludes the paper and outlines some future work.
\vspace*{-5mm}
\section{SDN Architecture and Threat Model} \label{sec:SATM}
\subsection{Basic SDN Architecture} \label{subsec:BSA}
\begin{figure}
    \centering
    \includegraphics[scale=.24]{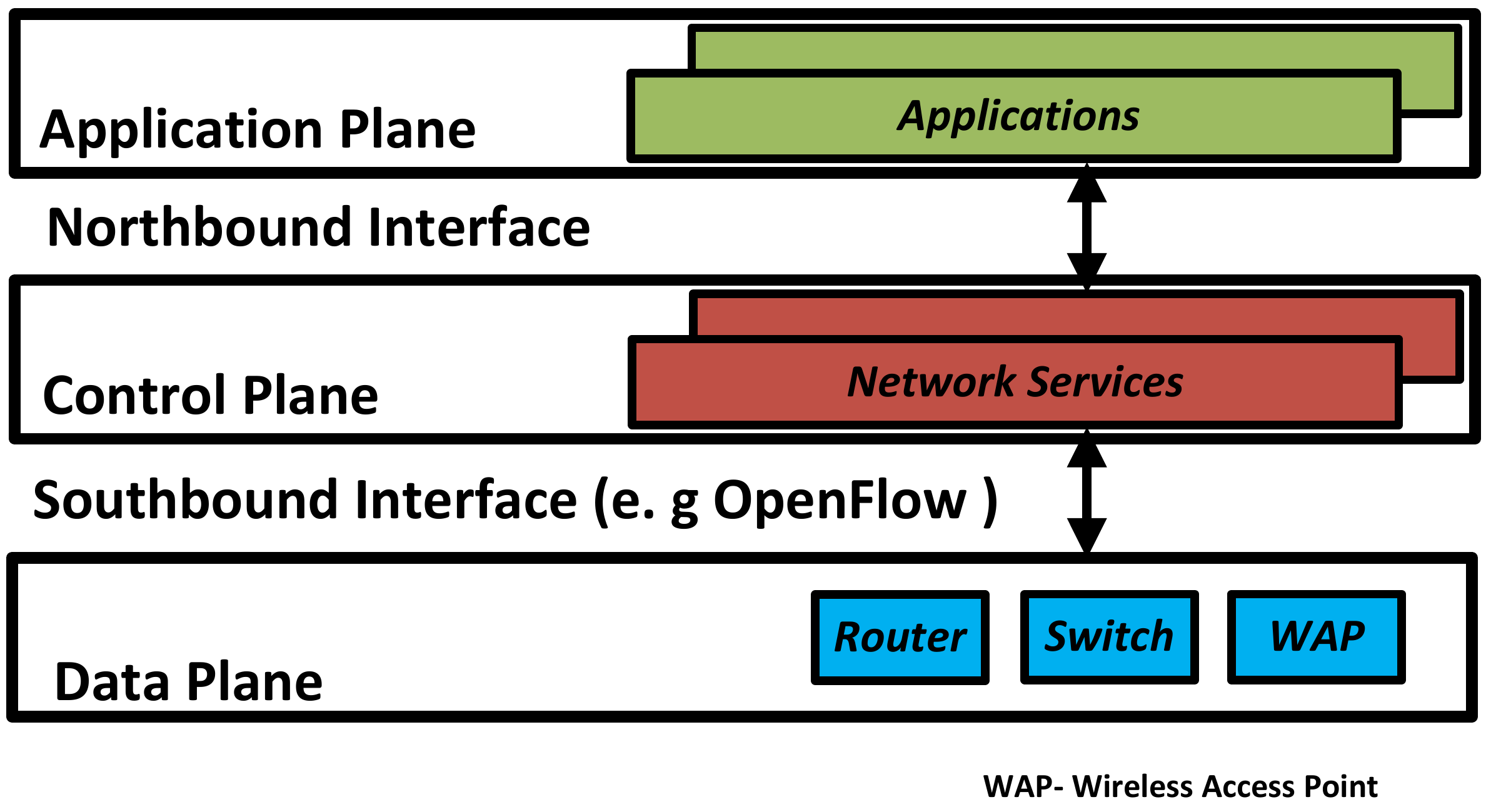}
    \caption{Standard SDN Architecture}
	\label{sdnArc}
	\vspace*{-3mm}
\end{figure}
\noindent Figure \ref{sdnArc} shows an overview of the SDN architecture with different components in the control plane and data plane. The control plane consists of a logically centralized Controller (which itself can be distributed in practice) with native applications for management of devices in a SDN network.  There can be several third party applications that can be hosted (or access different services) in the Controller. The interface between the Controller and the applications is referred to as the North Bound Interface. The data plane consists of the networking devices and the interface between the Controller and the networking devices is referred to as the South Bound Interface. OpenFlow \cite{mckeown2008}  is the most commonly used protocol for communication between the Controller and the networking devices. Other protocols include sFlow and SNMP that can be used for communication between the Controller and the network devices.   In the SDN terminology, a switch refers to any networking device that operates in Layers 2 to 7 in the OSI model. 

\noindent The Controller manages the flow-entries in the flow tables of the switches through a secure channel (that exists in the switch itself). This management process might be done both reactively (in response to packets) as well as proactively. There are several SDN Controllers \cite{kreutz15sur} available using different programming languages and environments. For example, NOX is based on C++ and Python programming languages; POX is based on Python; and Beacon and Floodlight are based on Java.

\noindent An OpenFlow compatible switch in the data plane contains three parts: i) A Flow Table, with an action linked with each flow entry, which tells the switch how to process the flow; ii) A channel that joins the switch to a remote control process (the Controller), allowing communication between a Controller and the switch;  communication instructions and packets to be sent between a Controller and the switch; using the OpenFlow protocol; 
iii) The OpenFlow protocol, which provides an open and standard way for a Controller to interconnect with a switch. 

\vspace*{-4mm}
\subsection{Our Network Architecture and Threat Model}
\noindent Our network architecture consists of multiple Autonomous System (AS) domains, with each domain having packet forwarding devices such as routers, gateways and switches and end hosts (which are connected to users). In general, there can be multiple SDN Controllers but for simplicity, we assume that each domain contains one SDN Controller.  End hosts are connected to the forwarding devices. For clarity, we will assume that each AS has a separate entry and exit gateway. These are OpenFlow supported forwarding devices. Though we assume OpenFlow switches, our solution will work with legacy switches, as long as there exists a path which has some OpenFlow switches. (Furthermore, migration solutions are being developed which allow incremental migration to software defined networks from current networks, e.g.~\cite{canini14software }). The traffic generated by the end hosts and forwarded by the network devices is subjected to security policies specified in the Controller. As there are multiple SDN domains, we have both intra-domain as well as inter-domain communications. In an intra-domain communication, the traffic from source to destination passes through devices within a single SDN domain and the requested services are provided by the servers and devices in the same domain. In this case, the traffic and service requests are subjected to security policies in the SDN Controller of that domain. Inter-domain communications involving multiple domains require cooperation between SDN Controllers, as the communications are subject to security policies in multiple SDN Controllers.

\vspace{0.03in}

\noindent There are different types of security attacks that are possible in a SDN. There can be attacks on the control plane communications between the switches and the Controller in a domain as well as attacks on the data plane communications. In general, threats in SDN domains can be categorized into: 

\begin{enumerate}
\item  Threats against a SDN Controller
\item  Threats against the networking devices (switches)
\item Threats against communications between the Controller and the networking devices:
\item  Threats against communications between different SDN Controllers in different AS domains
\end{enumerate}
\noindent 1. Threats against a SDN Controller: As the Controller is at the heart of a SDN domain, if the Controller is compromised, then the whole network under that SDN Controller is made vulnerable. A Controller can be compromised in three ways:

\vspace{0.03in}

\noindent i)	due to malicious software errors or bugs in the Controller software system such as the operating system; \\
\noindent ii)	due to threats arising from malicious or compromised applications running on top of or in the Controller; \\ 
\noindent iii)	due to threats from the underlying network devices such as the OpenFlow switches.

The threats and attacks arising in the first two categories (i) and (ii) are somewhat similar to those that occur in software and operating systems. Such attacks can include common ones such as cross-site scripting,  SQL injection, command injection and buffer overflow. For instance, if a SDN application uses a web-interface, then it can be used by an attacker to install a malicious script which can bypass the authentication mechanism in the Controller. This will enable the attacker to gain access to the Controller and carry put further attacks such as eavesdropping on the device traffic and delete/modify flows in the flow entry tables. Security techniques to counteract these threats and attacks are similar to those used in software system security. 
 These attacks are not the focus of this paper. Security mechanisms described in works such as \cite{lee2016smaller} and \cite{lee2017delta} discuss some such attacks in the Controller.

\vspace{0.03in}

\noindent Our security architecture addresses the threats against the Controller in (iii), which occur via the switching fabric. The connections that a switch can establish are based on the policies specified in our security architecture, thereby helping to detect attack flows and counteract them. For instance, malicious hosts connected to the switches can launch flooding attacks by forging IPs\cite{kreutz13}. These communication requests from the forged IPs will look like legitimate requests to the switches. In a proactive mode, they will be dropped immediately. In the reactive mode, since they are new flows, a switch will generate a request to the Controller. This can lead to denial of service attacks against the Controller. Our security architecture has fine grained policies which are able to detect malicious host and flooding attacks, and drop any further requests from this host by installing drop rules to discard further flooding from the switches connected to that host.

\vspace{0.03in}

\noindent 2. Threats against Network Devices: Forwarding network devices are usually physically closer to the attackers. Often it is the less sophisticated forwarding devices that are the main targets for attacks. Each device contains some number of flow rules in a flow table which are used to route the packets. For instance, an attacker who knows the IP range of the Controller domain can flood a network device with forged packets. Also, if an attacker can access multiple hosts, s/he can launch denial of service (DoS) attacks by continuously sending random IP packets. An attacker may also be able to listen on the links between any of the forwarding devices. This is possible because often there is no protection on packets or policies to route the packet through specific switches (thereby preventing access to the flows)\cite{benton2013}. This can also lead to subsequent attacks by spoofing the address of any of the hosts and launching DoS and Man-in-the-Middle attacks as well as illegal modification of the flow table entries. Our security architecture will help to detect attack traffic from malicious hosts towards the switches, and block them by dynamically installing policy rules in the switches.

\vspace{0.03in}

\noindent 3. Threats against the Communications between the Controller and the Network Devices: The protocol between the SDN Controller and the switches need to be protected against security attacks. Typically the Controller may establish a secure link with the forwarding devices using security protocols such as Transport Layer Security (TLS). Some Controllers may not use the TLS support option in the OpenFlow~\cite{benton2013}. In this case, it makes it vulnerable to traditional security attacks such as eavesdropping, unauthorized modification of traffic and masquerading. In this paper, we do not address confidentiality and related key management issues. The security architecture proposed in this paper deals with access and information flow attacks in SDNs. We are currently extending the security architecture with encryption mechanisms and associated key management authorities, thereby counteracting such attacks as well as for providing on demand protection of traffic. We mention this in the Conclusion. Hence we will not be addressing these communication threats in this paper. It is worth pointing out that the techniques used to achieve the confidentiality and key management services are somewhat traditional and well known. 

\vspace{0.03in}

\noindent 4. Threats against the Communications between SDN Controllers in different AS domains: In practice, large networks may require more than one SDN Controller per AS domain and there can be many AS domains. Hence there can be attacks on the communications between different SDN Controllers in a single AS or in multiple AS domains\cite{kreutz13}. Furthermore, there is a need to take into account legacy networks. Interception of traffic between Controllers can lead to many attacks such as information theft, spoofing and flooding attacks. Hence the need for a secure routing model for Controller to Controller communications.

\vspace{0.03in}

Our security architecture is concerned with authorized flows across multiple domains by enforcing security policies specified in different SDN Controllers. These policies detect attacks such as flooding and spoofing, and ensure only secure and authorized traffic flows occur between domains.  The focus of this paper is the design of configurable security policies that can enforce authorized information flows in a distributed SDN environment with multiple domains managed by different SDN Controllers.  Our architecture enables secure virtual partition of the network to achieve separation of flows and services, thereby reducing the attack surface in SDN. It addresses threats arising from malicious traffic generated by end hosts leading to attacks against switches and the Controllers; it detects and prevents unauthorized flows and unauthorized access to services in a distributed SDN. 

\vspace{0.03in}
\noindent \underline{Attack Scenarios}


In particular, our security architecture can deal with the following specific attack scenarios: 
\begin{itemize}

\item (A1) Unauthorized communications between end hosts within a domain as well as between end hosts residing in different domains: Attacks such as worms are often successful by exploiting the default permit communication between the hosts in current networks. The infected machine uses a random source address to scan for vulnerable machines and spreads the attack.

\item (A2) A malicious end host attempting to access services on the network for which it is not authorized: Insider threat is one of the difficult challenges when it comes to securing networks in organisations. Often this involves malicious users making use of their machines to access services for which they are not authorized.

\item (A3) Attacks originating from end hosts residing in certain end locations: Wirelsss LANs are often used in providing free access to any user. Attackers can misuse such free access networks for generating attacks. Similarly, there can be malicious ISP/AS domains that are used in the generation of attacks.

\item (A4) A malicious end host generating malicious traffic attacking the switches and/or flows through certain paths in the network: An attacker can generate malicious traffic that is destined to the victim switch and/or flood certain paths in the network with malicious flows.




\item (A5) Attacks that chain malicious flows coming from different locations at different times: A bot master who has control over several hundred compromised machines can generate attacks on the victim networks or AS domains using compromised machines from different locations at different times, with possibly different types of attack traffic.  Such situations can occur in inter-domain scenarios using attacks such as Crossfire \cite{kang13crossfire} and Coremelt \cite{studer09coremelt}. In Crossfire, a small set of bots directs low intensity flows to a large number of publicly accessible servers. These flows flood carefully chosen small set of links, effectively disconnecting the target servers from the Internet. In Coremelt, attackers only send traffic between each other to clog the network, rather than towards a victim host. It is difficult to deal with these attacks as the attack traffic appears as legitimate communications between the end hosts.  

\end{itemize}
We discuss the security analysis of these attack scenarios in section \ref{sec:SA} using our proposed architecture.\\
The deployment of our proposed security architecture will provide system managers of complex distributed networks with facilities to configure dynamically security policies which can detect and counteract different types of attack flows, and achieve secure provisioning of SDN services.

\vspace*{-4mm}
\subsection{Security Requirements}
This section discusses the security requirements that are addressed in the design of our policy based security architecture. 
\begin{itemize}

\item (R1): Security requirements can vary for different types of flows in a network. There can be security constraints on different flow parameters such as the time of flow, location of devices that are generating or receiving the flows, and delay and bandwidth requirements for the flows. For example, critical applications may have stringent requirements for the delivery of messages.  Sensitive applications may require that the traffic is transferred through a secure channel involving secure switches. Hence, there is a need to ensure that these specific flow requirements are provisioned correctly in a multi domain SDN environment. This requires the ability to handle flow specific security characteristics in inter domain communications in SDNs.    
\item (R2) There is a need to secure normal SDN operations, as the attackers can exploit the weaknesses in the SDN operations to generate different types of attacks. For example, currently Controllers do not validate the flow requests before establishing the routes to enable communication between the end hosts. For instance, the attacks such as the spread of worms in traditional networks can also happen in SDNs. A malicious host scan for random addresses to find vulnerable machines and spread the attacks, establishing routes to destination hosts.   
\item (R3) As a Controller has the visibility of its network domain topology and devices, there is a potential to develop secure northbound applications making use of the information available at the Controller for achieving end to end security within a domain. 
\item (R4) In a multi-domain situation, the end hosts are involved in communication with hosts connected to different networks. Hence, there is a need for to develop techniques for achieving end to end security over multiple domains. However as a Controller has visibility only over its domain, achieving secure inter domain communications requires secure co-operation between the Controllers in different domains.
\end{itemize}
We discuss how these requirements have been achieved by our security architecture in Section \ref{sec:SA}
\vspace*{-2mm}
\section{Policy based SDN Security Architecture}\label{sec:pbsa}

\subsection{Security Architecture Overview}
\noindent First, we will consider a Policy based Security Architecture for intra-domain interactions and then we will address the inter-domain communications enabling end-to-end SDN services across multiple domains. We will present a high level overview of the architecture and then describe each component in detail.

\vspace{0.03in}

\begin{figure}
    \centering
    \includegraphics[scale=.58]{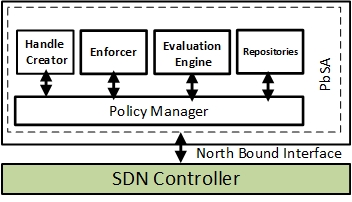}
    \caption{Policy based Security Architecture for SDN}
	\label{interArc}
	\vspace*{-3mm}
\end{figure}
\noindent Figure \ref{interArc} shows the Policy based Security Architecture for securing communication in a SDN domain. The Policy based Security Architecture can either form part of the SDN Controller or can run as a Security Application on top of the SDN Controller. We have designed and developed the Policy based Security Architecture as a Security Application running on top of a SDN Controller for flexibility reasons. We will refer to this as Policy based Security Application (PbSA). PbSA is implemented in the north bound interface of the Controller. As PbSA is designed in a modular fashion, the components of PbSA can be implemented on a single host or distributed over multiple hosts. 

\vspace{0.03in}

\noindent We assume that each AS domain is controlled by a SDN Controller. Each Controller has a Policy Server. The Policy Server has five main components the Topology and Policy Repositories, a Policy Manager, a Policy Evaluation Engine, a Policy Enforcer and a Handle Creator. Each Controller maintains and updates a Topology Repository and a Policy Repository. The Topology Repository contains the network topology information derived using traceroute mechanism mentioned below. The Policy Repository contains the Policy Expressions and specifications which are expressed using a simple language based template described in Section \ref{subsec:polrepospec} below. The Policy Manager as the name implies manages every single operation of the security system. An Evaluation Engine is used to evaluate the incoming network traffic against the relevant security policies for that specific traffic. Following the evaluation, the Policy Manager determines the flow rules which are then conveyed to the Enforcer module. The Enforcer module not only fetches the required information from the south bound interface connected to the switches but also enforces the flow rules obtained from the Policy Manager. A Packet Handle Creator module creates the necessary handles from the visited Controller which is piggy backed with the payload from the Policy Manager. These handles are used to check the authenticity of the packet and the enforcement of policies at the switches.

\vspace{0.03in}

\noindent With intra-domain communications, the traffic from source to destination passes through devices within a single SDN domain and the requested services are provided by the servers and devices in the same domain. In this case, the traffic and service requests are subjected to security policies in the PbSA in the SDN Controller of that domain. The routing process begins from the host that is generating the packets and the request, which is the source of the communication. This source host could be any client, such as a mobile device. The initial packet header from the source host is sent by the switch (to which this host is connected) to the SDN Controller in the AS domain. The header contains all the usual network and service parameters such as the source address, the packet type. The PbSA application in the Controller extracts the relevant parameters from the incoming packets and uses the Policy Repository and the Policy Manager to determine whether the relevant Policy Expressions are satisfied. If the Policy Expressions are valid for the incoming packets, then PbSA will enforce the specified actions as flow rules in the appropriate data plane devices such as switches to transfer the packets. 

\vspace{0.03in}

\noindent Inter-domain communications involving multiple domains require cooperation between SDN Controllers, as the communications are subject to security policies of multiple Controllers. Hence inter-domain routing of traffic requires an SDN Controller in one AS domain to have knowledge of other Controllers in other AS domains. To create a topological map of a distributed SDN environment, we have used the traceroute mechanism. Note that although there are other alternatives \cite{phan13collaborative} for topology discovery such as Border Gateway Protocol (BGP) and Internet Routing Registries, each one of them has its own issues. For the purpose of our prototype, traceroute was found to be sufficient and the easier one to use. It is not a critical part of our design and it is mainly an implementation issue. Each SDN Controller in an AS domain sends an ICMP signal using a different TTL level (1-6) to make a topological map of the AS domains. From the architecture point of view, each Controller has a Topology Repository to store the mapping of the topology information. SDN Controllers in each AS keep this Repository updated by running traceroute at different times. We have included an additional security attribute, a Security Label, in the ICMP response message from each AS SDN Controller. The intention of the Security Label is to reflect the level of security associated with that particular Controller. In our current architecture, this Security Label is hardwired (static) and is specified at the time of installation of the Controller based on the reputation of the manufacturer of the Controller. In the next stage of the development of the architecture, we will develop a meta-level security protocol that will enable secure and dynamic updating of the Security Label depending on the behaviour of the Controller over time. This will be done as part of a trust model which we are in the process of developing for distributed SDN environment. We have modified the ICMP response messages to attach the Security Labels. Hence each AS domain SDN Controller now has the ability to discover the topological information as well as the levels of security associated with all the neighboring AS domain Controllers in this Repository.

\vspace{0.03in}

\begin{figure}
\centering
\includegraphics[scale=0.32]{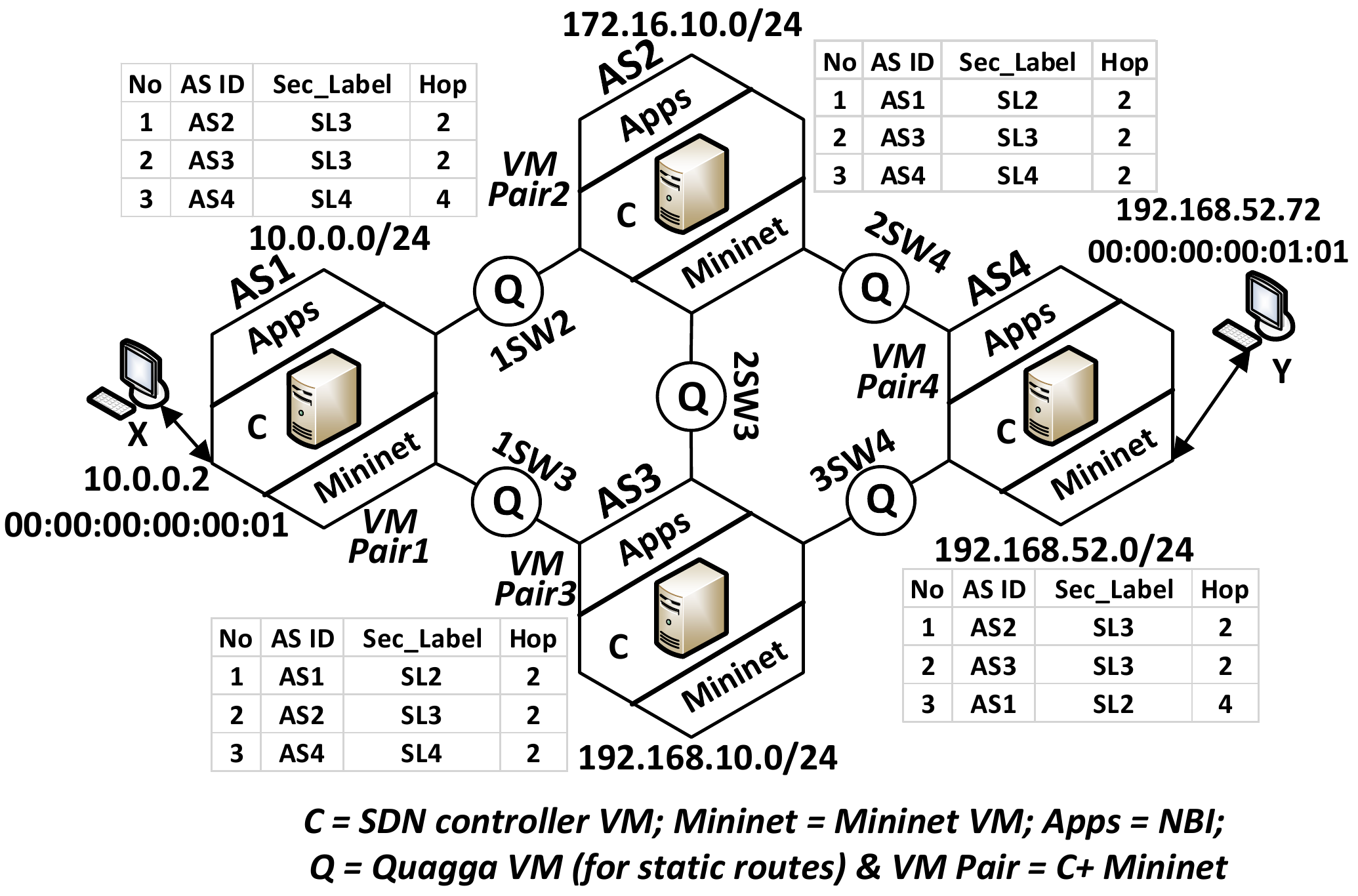}
\caption{Implemented Network with Routing Information.}
\label{interExample}
\vspace*{-3mm}
\end{figure}

Consider the distributed SDN environment shown in Figure \ref{interExample} and the associated Topology Repository tables are shown in the figure too. Each hexagon in Figure \ref{interExample} represents an AS domain. We have represented the AS Gateways using Gateway OpenFlow Switches. In this case, all the controllers use \textit{traceroute} by varying TTL level from (1-4). 
When TTL becomes 0, the Controllers responds with an ICMP TTL exceeded message, which contains additional information about the AS domain (Sender IP, ID, Security Label). This information is stored in the Topology Repository for future use by the respective SDN Controller. Each Table in Figure \ref{interExample} shows the Topological Repository for the respective domains SDN controller. \textit{ AS\_ID} is the identity of a particular AS, \textit{Sec\_Label} is the security label of the AS domain and Hops is the distance from source to destination AS. The edge OpenFlow Switches are represented using the notation ([source AS ID]SW[destination AS ID]), e. g. switch connecting AS1 to AS2 is represented as $1SW2$.

\vspace{0.03in}

In the inter-domain setting, our architecture introduces two additional mechanisms which have conceptual significance.  

\vspace{0.03in}

\noindent \underline{Handle}: The first mechanism is a {\em Handle}. PbSA creates a {\em Handle} and tags to each flow request. The Handle consists of a list of visited AS domain IDs. The {\em Packet + Handle} is then transferred to the next AS Domain Controller. A similar process is repeated as the packet goes through all the transit AS domain SDN Controllers until the packet reaches its destination. This Handle will be protected for integrity and it will be used in the validation of flows across multiple domains.  

\vspace{0.03in}
\noindent \underline{Policy Transfer Token}
\noindent The second mechanism is a {\em Policy Transfer Token}, which comprises policy constraints that are transferred from one AS domain to the subsequent transit AS domains and which need to be satisfied by the flow, as the packets are transferred. These constraints need to be taken into account in addition to the policy constraints of the transit domains. For instance, if there is a constraint that the traffic should only pass through AS domains with security label greater than a certain threshold, then this constraint needs to be satisfied by subsequent transit domains. Suppose an AS domain SDN Controller (with AS ID = 10) has a constraint that packets should only be forwarded through a path of AS domain SDN Controllers that have a security label greater than $SL3$.  In distributed systems, in general, it is not possible for one domain Controller to know about policies of other domains. Hence there is a need to transfer the policy constraints, which are communicated via \textit{Policy Transfer Tokens}. The significance of the transfer token is that the policy constraints that are transferred are only those policies that are specific to flows and packets in that flow. Such a mechanism is useful as it enables partial delegation of policies that are flow dependent. The next section describes in detail the specification of security policies. In this paper, we assume that the AS domain Controllers are secure and trusted, and that if and only if the policies can be satisfied in the domain, the receiving Controller will accept the packets. 
 
\vspace{0.03in}

\noindent In terms of the notation, we denote the Handle as $H_{i}^{AS_k}$ which is tagged to the packet (flow request), where $H_i$ is the Handle for a particular communication {\em i}, and {\em k} is the ID of the AS domain SDN Controller which created the Handle. Similarly, the \textit{Policy Transfer Token} is denoted as $PTT_{i}^{AS_k}$, where again $i$ denotes the specific communication and $k$ denotes the ID of the AS domain. Hence an AS domain SDN Controller creates the augmented Packet using the original Packet as well as the Handle and the \textit{Policy Transfer Token}.  

\vspace*{-3mm}
\subsection{Security Policy Specifications}\label{subsec:polrepospec}
\noindent A key component of the security architecture is the specification of security policies that are to be enforced on the SDN communications. The specified security policies are stored in a Policy Repository in the PbSA.

\noindent We have adopted a simple language based approach to specify the security policies. We have chosen the policy based routing syntax specified in RFC1102 \cite{clark89} as the basis for our security policy specifications. Policies are rules that specify a particular path or paths that packets must follow in the network and the conditions under which the packets follow these paths. In our language, we have Policy Expressions specifying a range of attributes associated with the flow and the entities in the SDN. 
These include the following:

{\fontsize{8}{9}\selectfont
\begin{itemize}
\item Flow Attributes: Flow ID, sequence of packets associated with the flow, type of packets, security profile indicating the set of security services associated with the packets in the flow;
\item Autonomous System Domain Attributes: AS identities such as source AS and destination AS identities (\textit{AS Domain ID}), sub-net address space (${SRC}_{SUB}$ for source and ${DST}_{SUB}$ for destination), identities of entry (${SRC}_{ENT}$) and exit (${DST}_{EXT}$) gateway/switch to AS, AS type (e.g. Commercial domain, Government domain) (${SRC}_{Type}$ and ${DST}_{Type}$ ) and security label associated with the AS (${SRC}_{SL}$ for source and ${DST}_{SL}$ for destination);
\item Switch Attributes: Identities of the switches and security label of the switches;
\item Host Attributes: Identities of hosts - source host IP \& MAC ( ${SRC}_{IP}$ \& ${SRC}_{MAC}$ ) and destination host IP \& MAC ( ${DST}_{IP}$ \& ${DST}_{MAC}$ );
\item Flow and Domain constraints: Flow constraints (\textit{FlowCons}) and Domain constraints (\textit{DomCons}) associated with a flow such as thresholds, attack signatures etc.;
\item Services - Services for which the Policy Expression applies;
\item Time Validity - The time period for which the Policy Expression is valid and
\item Path (${AS}_{SEQ}$) - In the case of intra-domain, indicates a specific sequence of switches whereas with inter-domain communications, it indicates the sequence of Autonomous Systems traversed by a flow. 
\end{itemize}
}
\noindent The flow constraints are conditions that apply to a specific flow or a set of flows. For instance, a constraint may specify that the flow of packets of a specific type (e.g video) should only go through a set of switches that can provide a certain bandwidth;  or from a security point of view, a constraint could be that a flow should only go through AS domains that are at a particular security level. Domain constraints apply to all flows within a domain. They are used to specify domain-wide policies. For instance, there could be a domain wide security policy which may specify that all flows should be protected for integrity, as part of the security profile. These constraints are used as part of the actions associated with the Policy Expressions. Packet ($PKT_{ATT}$) and Time attributes ($T^{PE_i}$) can be integrated into the constraints. Here, flow attributes indicates attributes associated with the sequence of packets in the flow such as type of the incoming packets based on port numbers, thresholds, security services associated with the packets and attack signatures. Time attributes represents the duration time for which a particular Policy Expression is valid. 

\vspace{0.03in}

\noindent Alternatively, it is also possible to enforce specific paths by explicitly specifying the set of switches through which a flow must go through or a specific set of AS domains that should be traversed. 

\vspace{0.03in}

\noindent The policy language has wild cards in its syntax enabling specification of policies that can apply to sets or groups of entities and services. When a Policy Expression is satisfied, then the associated action is performed which could be simple as allow or deny the request. Hence using these policy terms, one is able to specify different sets of Policy Expressions to deal with a range of scenarios in both intra-domain and inter-domain communication in a distributed SDN. An action can also have some attributes. For instance, destination exit switch (${DST}_{EXT}$) attribute associated with an action indicates the exit switch through which a flow should pass, once a policy is satisfied. Hence we are able to specify conditional policies, constraint and state dependent policies as well as obligation policies.

\vspace{0.03in}

In particular, the language can be used to specify policies that take into account the context associated with the resources and the devices. For instance, it can be used to specify protection policies that take into account the attributes of the devices through which the  flow can occur or be displayed. For certain confidential information, the paths through which the packets are transferred and the devices/switches which can process them must have certain security attributes. 
Such policy expressions can be specified using simple Boolean algebra on the security labels. The policy engine evaluates the Boolean expression to determine whether the condition on the security labels is satisfied or not.
The language can also be used to specify release policies associated with the end points through which the traffic can be released, requiring certain security attributes. These types of policies, namely protection and release policies, are common in the context of content based security, and are significant when it comes to the provision of SDN services.

\vspace{0.03in}

\noindent Using the policy terms mentioned above, a simplified Policy Expression template could be as follows:\\
\begin{small}
\noindent $PE^{AS_{k}}_{i}$= $<Flow ID, Source AS, Dest AS, Source Host IP,$\\ \indent$ Dest Host IP, Source MAC,
Dest MAC, User, FlowCons,$\\ \indent$ DomCons, Services, Sec-Profile, Path>:<Actions> \inlineeqno$
\end{small}

where $i$ is the Policy Expression number and $k$ is the AS ID. This is a generic Policy Expression for both intra and inter-domain. In the following sections, we will explain the use cases for both intra and inter-domain. For simplicity, we have omitted the AS ID notation in intra-domain expressions. Also, in intra-domain, \textit{path} indicates a set of switches within the domain while in inter-domain \textit{path} refers to a set of AS domains. \\
In general, we have a number of Policy Expressions stored in the SDN Policy Repository. Such a template enables us to specify a range of policies for different users (and hosts), from different locations, accessing different services using different devices following different paths. Later we will illustrate the use of such policy expressions in both intra and inter domain environments when we discuss the results of our system implementation in Section \ref{ssec:ie}

\vspace*{-4mm}

\subsection{Security Architecture Walk-through}\label{sec:SAOWT}
\noindent Let us now give a brief walk-through of the operation of the proposed security architecture described above. We will use the inter-domain scenario given in Figure \ref{interExample} to illustrate the various steps involved. The Tables beside each of the AS domains in Figure \ref{interExample} represent the Topology Repository. Table \ref{polTerm} shows the policies in each of these AS domains stored in the Policy Repositories of PbSA.

\vspace{0.03in}

\noindent The initial packet header from the source host is sent by the switch (to which this host is connected) to the SDN Controller in that AS domain. The PbSA application in the Controller extracts the relevant parameters from the incoming packets and uses the Policy Repository and the Policy Manager to determine whether the relevant Policy Expressions are satisfied. If the Policy Expressions are valid for the incoming packets, then PbSA will enforce the specified actions as flow rules in the appropriate data plane devices such as switches to transfer the packets. 

\begin{table}[]
\centering
\caption{Stored Policy Terms}
\label{polTerm}
{\tiny
\begin{tabular}{cc} 

\hline
AS ID & Policy Expression \\ \hline
AS1   & \pbox{20cm}{$ PE_{1}^{AS1}=<*, (10.0.0.0/24, EDU, SL2), (192.168.52.0/24), 10.0.0.2, $\\ $ *, *, *, *, *, SL2+=, (80,443), conf, *>:<(1SW2, Allow )> $}
\\
AS2   & \pbox{20cm}{$ PE_{4}^{AS2}=<*, (10.0.0.0/24, EDU, SL2), (192.168.52.0/24), 10.0.0.2, $\\ $ *, *, *, *, *, SL2+=, (80,443), conf, (AS1)>:<Allow>$}
\\
AS3   & \pbox{20cm}{$ PE_{2}^{AS3}=<*, (10.0.0.0/25, EDU, SL2), (192.168.52.0/24), $\\ $ *, *, *, *, *, *, SL2+=, (80,443), conf, (AS1,AS2)>:<Allow>$}
\\
AS4   & \pbox{20cm}{$ PE_{2}^{AS4}=<*, (10.0.0.0/25, EDU, SL2), (192.168.52.0/24),$\\ $ 10.0.0.2, 192.168.52.72, *, *, *, *, *, (80,443), conf,$\\ $ (AS1,AS2,AS3)>:<Allow> $}
\\
\hline
\end{tabular}}
\end{table}

\noindent In this scenario, the host machine $X$ (with IP address $10.0.0.2$) wishes to communicate with the host machine $Y$ (with IP address $192.168.52.72$). As $X$ and $Y$ reside in two different AS domains, communication between them occurs via transit AS domains. At first, packets from $X$ go to the SDN Controller of AS1. As the Policy Expression $PE^{AS1}_1$ in AS1 matches with this particular network traffic, HTTP \& HTTPS traffic originating from $10.0.0.2$ are allowed to go to $Y$ in the subnet $192.168.52.0/24$. However let us assume that there is a flow constraint which specifies communications between $X$ and $Y$ must occur only through domains which have security levels greater than or equal to $SL2$. This will result in the traffic routed through the AS2 domain via the OpenFlow Switch \textit{1SW2} (Connected to AS1). A similar process occurs in AS2 and the traffic is sent to $Y$ in AS4.

\vspace{0.03in}

We also have constraints such as a flow or flows between two entities in a domain A and B should only go through paths whose security labels are greater than or equal to a specific security label L. Satisfying this rule requires the PbSA to determine the labels of the various paths between the devices, and then check whether the flow constraint is satisfied. The policy rules can have Boolean expressions such as the security label of the switch $S_i$ should be less than or equal to the security label of the neighbouring switch $S_{i+1}$. If the neighbouring switch meets this constraint, then the flow will be directed through it; if not, than another neighbouring switch will be selected.  In general, flow and domain constraints can require some form of evaluation which is more than just policy matching.

\vspace{0.03in}

\noindent In general, our architecture can be used in either reactive or in a proactive mode. In the reactive mode, the first packet of the flow received by switch will trigger the insertion of flow entries in switches in the network. This approach presents the most efficient use of existing flow table memory, but every new flow incurs an additional setup time. In the proactive approach, the flow tables in the switches are pre-configured by the Controller based on policies specified by the network administrators. This approach has no additional flow setup time because the forward rule is defined.

\section{Security Architecture Implementation} \label{sec:SAI}
\noindent \noindent We have implemented and validated our Policy based Security Architecture for SDN using Open Network Operating System (ONOS). We have modified the SDN-IP application in ONOS, and added extra modules for policy control for SDN inter-domain communications. We have used the Oracle VM Box environment to create ONOS and Mininet VM pairs, which act as inter-domains. Each inter-domain SDN Controller runs our PbSA over ONOS. 

\vspace{-3mm}

\subsection{Implementation Components}

\noindent {\bf Topology Repository:} Topology is stored in the Topology Repository and is implemented using the SDN-IP application. The repository contains topological information of the neighboring domains as well as the Security Labels associated with the SDN Controllers in these domains. As mentioned earlier, in the current version of the security architecture, these Security Labels are statically assigned based on the manufacturer of the SDN Controller and the reputation of the manufacturer. However in the next version of the security architecture design, we intend to update these Security Labels using a dynamic trust model for SDN. 

\vspace{0.03in}

\noindent {\bf Policy Repository:} Policies are represented as Policy Expressions in our template based language specifications and stored in the JSON Policy Repository. We have used the Java parser to update the JSON Policy Repository. Collections of Policy Expressions specify the various policies as mentioned in Section \ref{sec:pbsa} and they control the interactions and packet transfers both within and between AS domains. 

\vspace{0.05in}

\noindent {\bf Policy Manager:} Coordinates the AS policy routing system and is responsible for the following major tasks:

{\fontsize{8}{9}\selectfont
    \begin{itemize}
        \item Receiving information from the edge OpenFlow switches via the Policy Enforcer module.
        \item Aggregating the necessary information to match the Policy Expressions.
        \item Separating the handles from the payloads in the packets.
        \item After processing by the Policy Evaluation Engine module, sending the information to the Packet Handle Creator to create the new handle with the previous payloads.
        \item Creating the new packet using the newly created handle and payload.
        \item Sending instructions to the Policy Enforcer to install the respective flows.
    \end{itemize}
}
\vspace{0.03in}

\noindent {\bf Policy Evaluation Engine:} This module is used for determining the policies related to the flow. Relevant parameters are extracted from south-bound traffic with the help of Policy Enforcer module. The Policy Manager sends these parameters to the Policy Evaluation Engine.  The Engine module checks these parameters against the Policy Expressions stored in the Policy Repository. When an appropriate match for the entry is found, it enforces the corresponding Policy Expression(s) using the Policy Enforcer module via the Policy Manager. Creation of the Topology Repository and sending information to/from the Topology Repository to the Policy Manager are also performed by the Policy Evaluation Engine. The implementation section below gives the schema of the JSON based Policy Repository. 

\vspace{0.03in}

\noindent {\bf Policy Enforcer:} It enforces the flow rules as specified by the Policy Manager for the specific incoming south-bound traffic. 
\vspace*{-5mm}
\subsection{Network Setup}
We have used ONOS as the SDN Controller in each AS domain. The machine we are using for simulation purposes is a Core i7 - 4790 with 3.60 GHz CPU and 32 GB of RAM. Each SDN controlled AS is constructed by a pair of ONOS based VM and mininet VM. 
Our network configuration is shown in Figure~\ref{interExample}. We have added bridged mode Ethernet adapters to the VM pairs within the VM Box to create communication channels. BGP route advertisement is done by QUAGGA~\cite{ishiguro2007}, which is a BGP software router. This experimental network setup follows the same approach as the one taken by the ONOS technical team for their demonstration of SDN-IP application~\cite{jonathanhart14}. 


\subsubsection{Application Modules}
\begin{figure}
\centering
\includegraphics[scale=0.30]{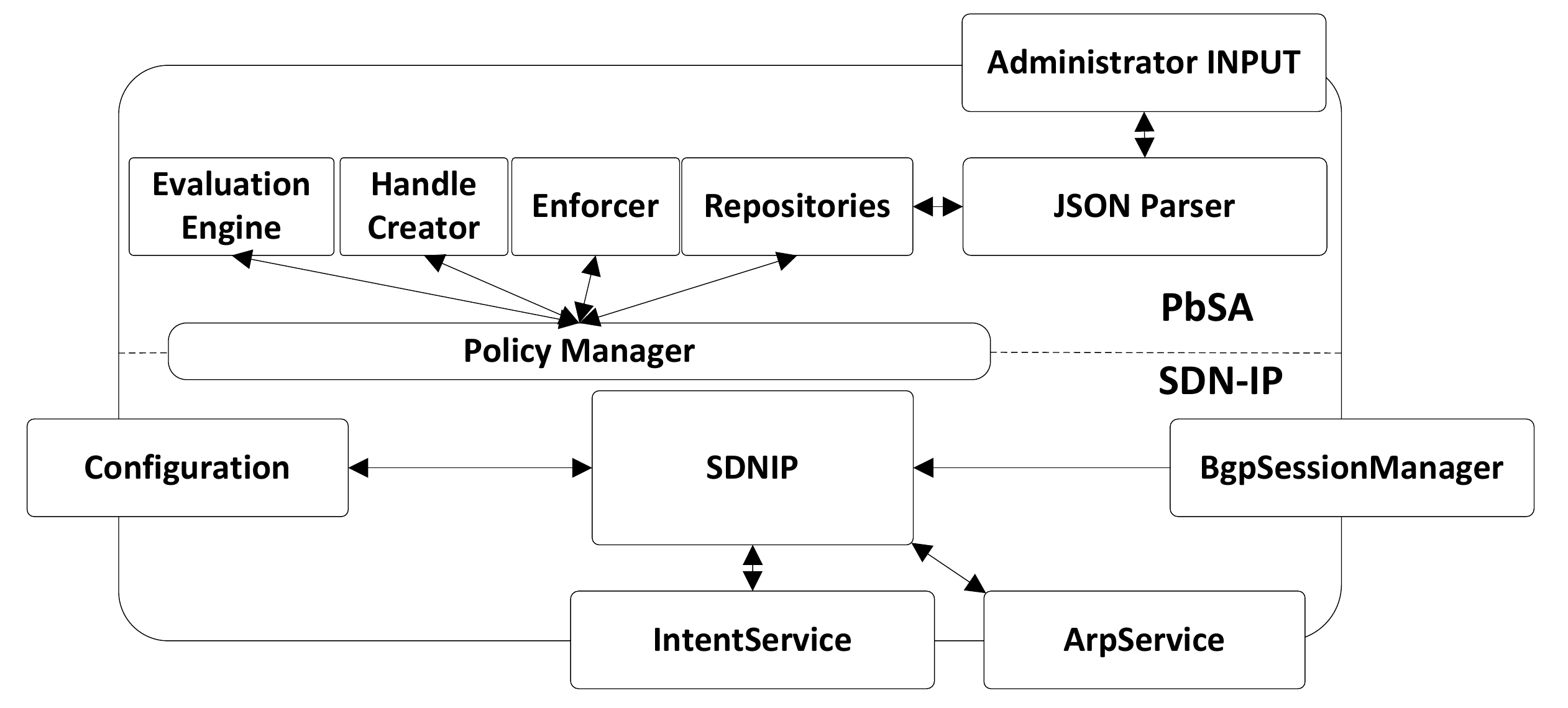}
\caption{PbSA Software Modules}
\label{ap}
\vspace*{-3mm}
\end{figure}    
\noindent Figure ~\ref{ap} shows the different modules used in the construction of PbSA. We have combined our application with SDN-IP (ONOS built-in application for BGP communications). Here we use SDN-IP BGP route selection and update for maintaining session modules. PbSA uses this information to create and update the topology in the Topology Repository. 

\noindent SDN-IP core consists of a software router which captures the best BGP routes from BgpSessionManager. BgpSessionManager maintains sessions, updates and selects the best BGP routes. ArpService is used for MAC address resolution. ONOS core configurations are captured by the Configuration module. IntentService submits the requests in the form of intents to the ONOS controller. Using these modules, SDN-IP listens to BGP requests and chooses the best route. SDN-IP has certain limitations which are mentioned in the related work section (Section \ref{sec:RW}). 

\noindent As mentioned above, JSON Repository is used for storing the Policy Expressions. To update, modify and read the stored JSON policies, we have used JAVA JSON parser module (a built-in module in Java). The Policy Manager is a core component of the PbSA. It maintains all the modules in the PbSA. The Policy Enforcer acts as a bridging module between the SDN-IP and our application. It extracts the topological information from SDN-IP and updates the topological database. The Policy Manager is aware of the topological information via this database. Before setting up any new flows between the available domains, the Policy Manager checks the JSON policy repository using Policy Evaluation Engine. The Policy Evaluation Engine parses the particular Policy Expressions and provides them to the Policy Manager. It compares the policies and sends the particular action to the Policy Enforcer, which forwards it to the intent services. These intent requests are captured by the ONOS Controller and appropriate action is taken. The Handle Creator is used to create the packets.  
\vspace*{-1mm}
\subsubsection{Policy Example in the Database}
{\fontsize{7}{8}\selectfont
\begin{lstlisting}[caption=AS2 ONOS Policy Database Sample ,label=poldb]
[{
  "id": "21",
  "flowid": "*",
  "srcasid": "*",
  "srcassub": "10.0.0.0/25",
  "srcastype": "EDU",
  "srcastrulabel": "SL2",
  "dstasid": "*",
  "dstassub": "192.168.52.0/24",
  "dstastype": "EDU",
  "dstastrulabel": "SL4",
  "srcip": "10.0.0.2",
  "dstip": "192.168.52.72",
  "srcmac": "00:00:00:00:00:01",
  "dstmac": "00:00:00:00:01:01",
  "user": "",
  "flowcons": "*",
  "domcons": "SL2+=",
  "services": "*",
  "secprof": "conf",
  "seq": "AS1, AS2",
  "action": "allow"
   }]
\end{lstlisting}
}

\noindent Listing ~\ref{poldb} shows a single Policy Expression from AS2 (VM pair 2) Policy Repository (JSON file). This single Policy Expression (ID 21) states that, for any packets originating from the subnet $10.0.0.0$, host $10.0.0.2$ with the MAC address ending in $:01$, whose destination subnet is $192.168.52.0$ must be routed through the domains whose Security Label is greater than or equal to $SL2$.

\vspace*{-1mm}

\section{Results}\label{sec:R}
\subsection{Implementation Examples}\label{ssec:ie}

\noindent 
This section presents specific implementation intra and inter domain scenarios using our architecture. 



\subsubsection{\bf Intra-domain Scenario}


\noindent \textit{Path Based Policy for Services}: 
\begin{figure}
\centering
\includegraphics[scale=0.36]{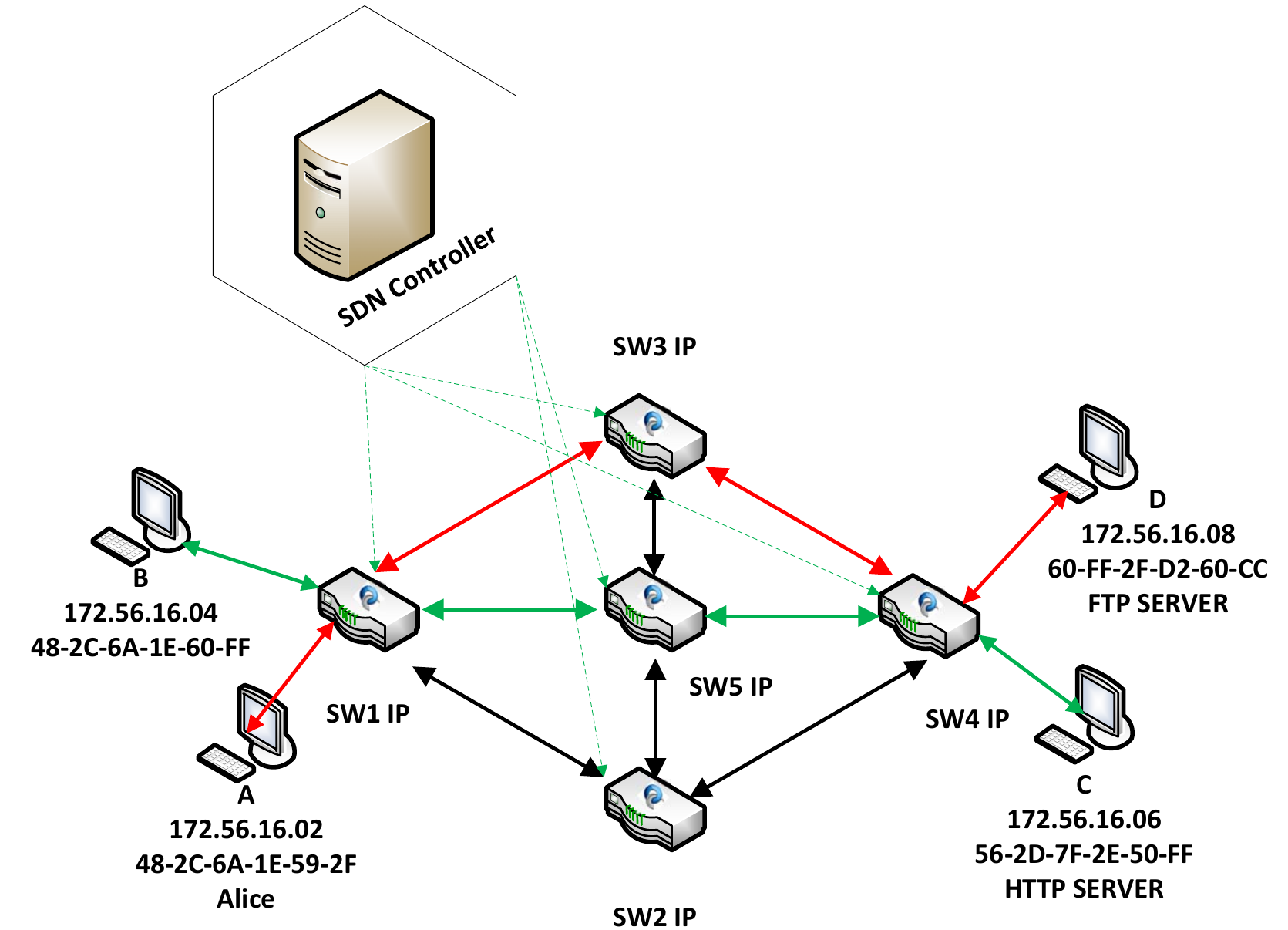}
\caption{Intra-Domain Example Scenario}
\label{intrasce3}
\vspace*{-3mm}
\end{figure}
This scenario considers specific paths for specific services requested from the hosts. Consider, for example, HTTP and FTP servers running on two machines with IP addresses $172.56.16.06$ and $172.56.16.08$ respectively. We have used Oracle VM box for this purpose. We have created an OpenFlow switch matrix with five switches as shown in Figure \ref{intrasce3}. Flows from two hosts with IP addresses $172.56.16.02$ and $172.56.16.04$ are guided through this switch matrix depending on the services they are intending to access. The Policy Expressions associated with these requests are as follows:
{\fontsize{8}{8}\selectfont
\noindent $PE_{3}$=$<*, *, *, 172.56.16.04, 172.56.16.06, 48:2C:6A:1E:60:FF, *, *, *, *, 80, \{Conf, Intg\} , (SW1;SW5;SW4)>:<Allow> \inlineeqno  $\\
\vspace{0.03in}
\noindent $PE_{4}$=$<*, *, *, 172.56.16.02, 172.56.16.08, 48:2C:6A:1E:59:2F, *, *, *, *, (20;21;22;23), Conf,(SW1;SW3;SW4)>:<Allow> \inlineeqno $}

\vspace{0.03in}

\noindent Equation 2 ($PE_{3}$) states that flows from host with IP $172.56.16.04$ and MAC $48:2C:6A:1E:60:FF$ accessing the HTTP server on IP $172.56.16.06$,
should be protected for confidentiality and integrity, and forwarded via OpenFlow switch SW1-$>$SW5-$>$SW4.\\
\noindent According to the Equation 3 ($PE_{4}$), flows to the FTP server running on IP $172.56.16.08$ from host with the IP address $172.56.16.02$ and MAC address $48:2C:6A:1E:59:2F$ are to be forwarded via a OpenFlow switch path SW1-$>$SW3-$>$SW4. This Policy Expression also instructs the Controller to protect the flow for confidentiality.
\begin{figure}
\centering
\includegraphics[scale=0.38]{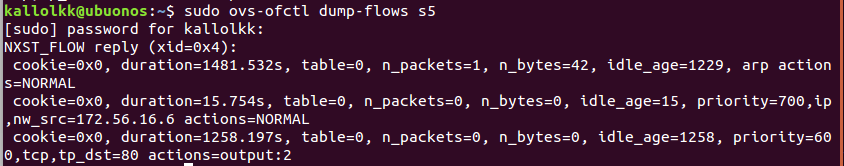}
\caption{Flowdump of the Switch SW5}
\vspace*{-3mm}
\label{fd}
\end{figure}
Also, all HTTP packets are to be routed through SW1-$>$SW5-$>$SW4. 
\vspace{0.03in}
Figure ~\ref{fd} shows the flow dump of the SW5 switch. The first flow is used for network discovery (ARP). The second flow is for the response from the Web Server running in IP $172.56.16.06$, which tells the Controller that any IP packet from the Web Server should be handled normally. The third flow is dedicated for HTTP packets and indicates that all HTTP traffic should take an output path Port 2 (in the Open vSwitch). For the FTP traffic towards the FTP server, similar type of flows are being installed in Switches 3, 1 and 4. But this time FTP traffic is used in the flow rules instead of HTTP.
\subsubsection{\bf Inter-domain Scenario}
\noindent {\em Policy specific to User, MAC, IP Addresses:} In this scenario, specific users, devices and IP addresses from one AS domain are allowed access to other AS domains running our application over ONOS Controller. For instance, this scenario can correspond to users from one organization visiting another organization and trying to communicate with their home organization corresponding to a roaming BYOD. This scenario is shown in Figure ~\ref{interExample}. Here, user Alice is an employee of an organization belonging to domain $AS2$, which is a restricted domain for only authorized users. We assume that her device's MAC address and the user ID are stored in a database, and that our policy application uses this policy database for checking user Alice's validity.\\
\noindent Let us now consider Alice roaming with her mobile device using company X's ($AS1$) Internet service. Alice requests to get access to HTTP employee server running in AS2. Since there is no policy permitting this flow to the restricted domain, it will be dropped at the edge router of $AS2$ domain. 

\begin{small}
\begin{math}
PE_8^{AS2}=<*,*,AS2,*,(172.16.10.66), (79:c8:82:b2:7b:1a),*, Alice, *, *, (80,443),* , * >:<allow>
\inlineeqno
\end{math}
\end{small}

\noindent Let us now introduce the Policy Expression 8 ($PE_8^{AS2}$). This Policy Expression tells the SDN Controller that any traffic specific to Alice's device ($MAC:79:c8:82:b2:7b:1a$) for HTTP traffic must be allowed in AS2 domain. In this case, when the packet comes to the edge OpenFlow router, it does not have any flow rule to guide the traffic; so it forwards the traffic to the SDN Controller. In PbSA, the Policy Evaluation Engine checks the packet information against the policies in the Policy Repository. It finds a match with the Policy Expression  $PE_8^{AS2}$. Therefore, the Policy Manager asks the Policy Enforcer to install the flow rules in the specific switches. Finally, Alice is able to communicate with the employee web portal running in the AS2 domain.  

\vspace{0.1in}

\subsubsection{\bf Inter-domain Scenario} {\em Policy based Routing for Unknown Communications:}  
Restricted AS domains often get requests from unknown domains to pass their traffic. Passing such traffic can often lead to threats in the transit domains. PbSA has a default deny policy for dropping all the flows without a matching permit policy. However, there is an option in the PbSA to permit such unknown flows through switches with low security labels. This will enable separation of secure flows from unknown flows.  Hence even if unknown traffic leads to attacks in the network, this will not impact the sensitive flows in the switches with high security labels. Such policies help to create secure channels in the network dynamically based on policies. Consider the following: \\ 
\begin{small}
\begin{math}
PE_{10}^{AS2}=< *, ( 10.0.0.0/25, EDU, SL1), *, *,\\ *, *, *, *, *, SL1,(80,443), *, *>:<allow>
\inlineeqno
\end{math}
\\
\begin{math}
PE_{14}^{AS3}=< *, (10.0.0.0/25, EDU, SL1), *, *,\\ *, *, *, *, *, SL1, (80,443), *, ( AS1, AS2)>:<allow>
\inlineeqno
\end{math}
\end{small}

Here $AS2$ and $AS3$ are transit domains. Within a domain, recall that the edge routers and switches are tagged with lowest security labels. In this scenario, $SL1$ represents the lowest security label. A Web Server is running in $AS4$. Some host in $AS1$ is trying to communicate with the Web Server. According to the topology repository, there is no direct passage to $AS4$ from $AS1$.  $AS1$’s packets need to go via transit domains $AS2$ and $AS3$. Assume that the domains $AS2$ and $AS3$ are restricted and commercial domains, and hence access through these domains is limited and controlled by the PbSA policies.
Policy Expressions $PE_{10}^{AS2}$ and $PE_{14}^{AS3}$ are stored in the Policy Repositories of $AS2$ and $AS3$ domains. The first Policy Expression $PE_{10}^{AS2}$ states that, any packet originating from domain $AS1$ $(10.0.0.0/25)$  and destined to Web Server in $AS4$ without a matching policy should  be  given  passage  through switches with lowest security label $(SL1)$ in $AS2$.  Hence  in  this domain, specific transit switches bearing a security label $SL1$ are  used  to  route  the  traffic  for  this  particular  host.  Similar situation occurs with the $AS3$ transit domain. Hence the specific host can access the Web Server in $AS4$.


\subsubsection{Flooding Attacks from Malicious Hosts}\label{subsec:fld}
In this case, an adversary connected to one of the OpenFlow switches in our network shown in Figure \ref{intrasce3} carries out a flooding attack. Assume that the prime motivation of the adversary is to exhaust the Controller by overflowing it with Packet\_in requests.  
In our network testbed, this is achieved by replacing Alice's VM with a Kali Linux VM and launching for instance a SYN flooding attack. Without PbSA, all the SYN messages resulted in the establishment of routes.  When the PbSA was activated, it detects requests above the specified threshold from a switch, and installs a block rule in the switch to drop the traffic from the malicious host carrying out the attack.  In our architecture, we have used two techniques to counteract such attacks. The first one involves a heuristic rule based mechanism using thresholds. This enables us to determine a threshold for each host connected to the switch and a threshold for each switch connected to the Controller.  These thresholds are dependent on the number of hosts connected to a switch and the number of switches connected to the Controller respectively. Here is a simple example that illustrates these thresholds.  Let 'CC' be the total capacity of Controller that is connected to 'X' switches. Let 'CS' be the capacity of the switch and 'Y' is the number of hosts connected to each switch. Then threshold 'TS' for each switch is determined by TSw = CC/X and threshold for each end host is determined by Thost = TS/Y. For example, consider the case of a single Controller that is connected to 10 switches which are further connected to 10 end hosts. If the Controller is capable of processing 1000 requests per  second then each switch is restricted to sending 1000/10 = 100 requests per second and each end host is restricted to sending 100/10=10 requests per second. We dynamically vary the weights associated with the calculation of threshold based on history of packets received; the threshold policy also varies depending on the number of running instances of the Controller. For instance, the Controller services can be hosted on multiple machines during peak hours. As a second technique, we have experimented with machine learning based technique to detect anomalies. In this case, the history of packets received is used with a machine learning algorithm to detect whether an attack is happening. We have used Bayesian algorithm for this purpose. We are currently looking into extending this aproach using other machine learning algorithms \cite{preeti18sur}.\\
Furthermore, note that with our PbSA architecture, as each flow request is validated by the Controllers in the domains before permitting the flows, attacks such as Crossfire \cite{kang13crossfire} and Coremelt \cite{studer09coremelt} are prevented at the source, which is more efficient. Even if the attacker generates untrusted traffic with unknown applications to create these attacks, policies such as those in Equations 5 and 6 help to route all the unknown and untrusted domain traffic via the switches with lower security labels. PbSA uses policies to compartmentalize the forwarding layer based on the security labels of the switches, thereby protecting the trusted traffic from unknown traffic.
\vspace*{-3mm}
\subsection{Performance Results}
We have conducted extensive analysis of the proposed architecture. In this paper, we only present a subset of these results due to space restrictions.  In particular, we will present the results related to the secure establishment of routes in intra domain and inter domain with the PbSA that we have discussed above.  The results vary for different types of hardware and the configuration of the devices. Hence, first we present the baseline result, which shows the performance of the Controller without the PbSA.  Then we enable the PbSA on the Controller with the same hardware and baseline settings to obtain the performance results specific to PbSA. \\
We first present results for latency, flow establishment rate, and memory overhead in intra-domain scenarios. Then we present the results related to the inter-domain scenarios.  
\begin{figure}
\centering
\includegraphics[scale=0.30]{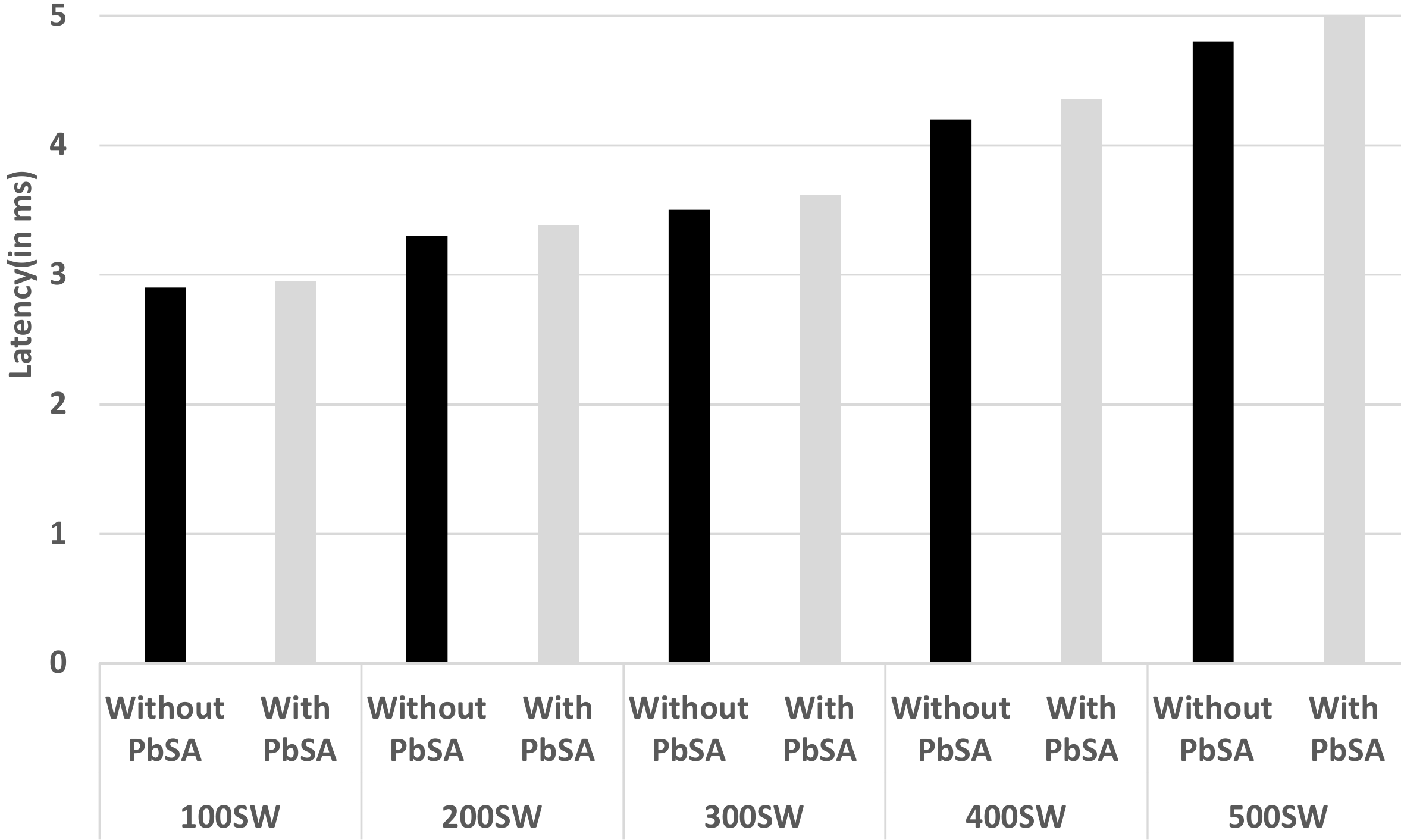}
\caption{Latency at the Controller}
\vspace*{-3mm}
\label{fig:lvss}
\end{figure}
The latency is the time taken by the Controller to process the Packet\_in message and respond with the Packet\_Out message. It is measured by sending only one Packet\_in message to the Controller without enabling PbSA to obtain the baseline results for the Controller. Then we enable PbSA with the similar setup to obtain the latency results. The process is repeated 10 times for each scenario and then we present the average result for each case. We repeat this process with varying number of switches. The results show that there is an increase in the latency with the PbSA compared to the baseline (without the PbSA). Also note that the latency increases linearly with the increase in the number of switches with the PbSA.  
\begin{figure}
\centering
\includegraphics[scale=0.29]{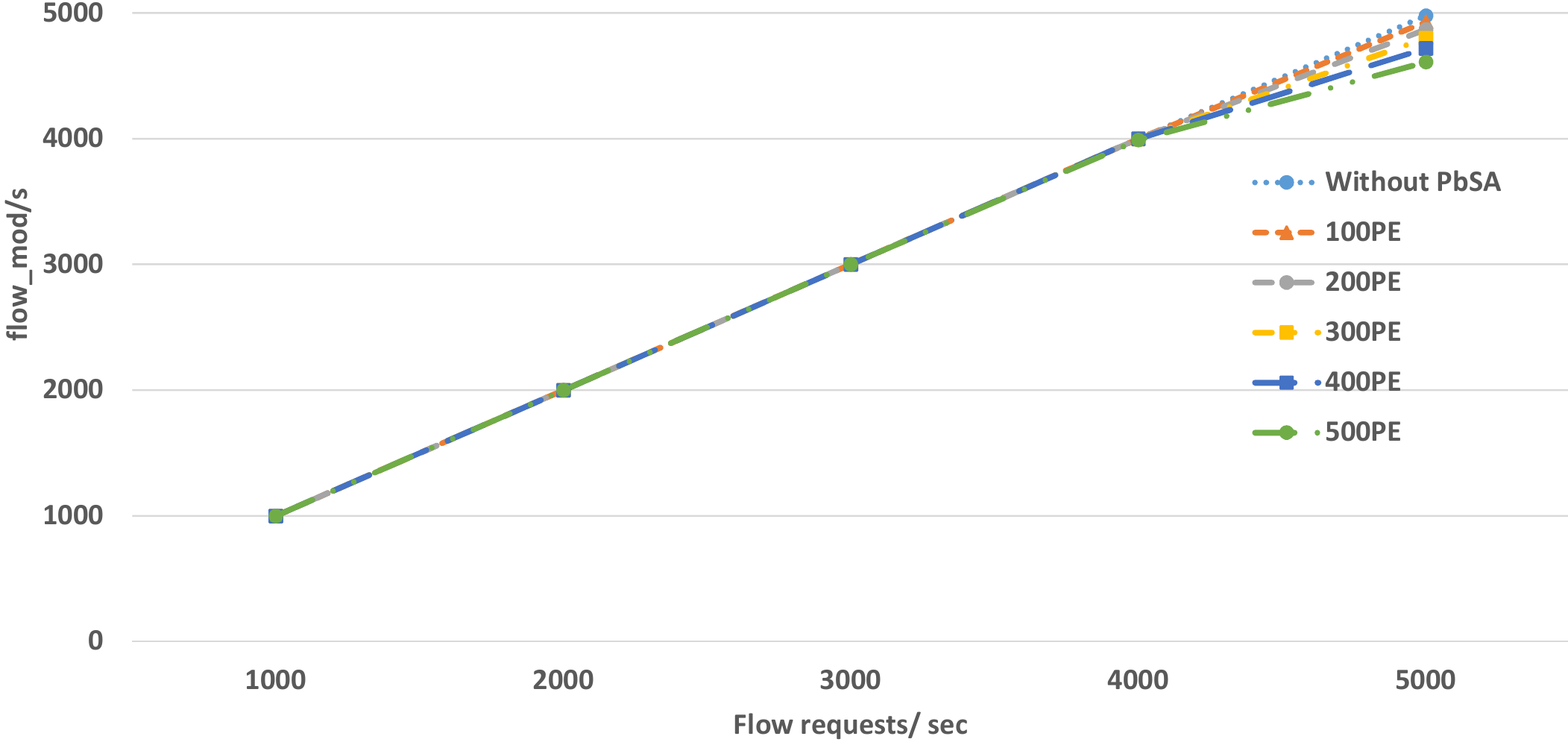}
\caption{Flow establishment rate of the Controller}
\vspace*{-3mm}
\label{fig:ferc}
\end{figure}
Flow establishment rate is measured by analysing the number of flow request messages that can be processed by the Controller per unit time with $500$ switches in the network.  We have tested the flow establishment rate by increasing the number of flow request messages and observing the number of flow\_mod messages generated by the Controller in the baseline mode and then compared with PbSA, while varying the number of policy expresssions (PEs). As shown in Figure \ref{fig:ferc}, there was $100$ percent flow establishment in the baseline mode and with PbSA, when the flow requests are generated at the rate of $4000$ requests per second. However, we see a variation in the flow establishment when the flow requests are generated at $5000$ requests per second. In this case, the baseline mode was successful in generating $4980$ flow\_mod messages and there was a variation in the number of flow\_mod messages with PbSA. The results varied with the number of PEs stored in the PbSA. We have observed flow establishment rate of $4934$ with $100 $ PEs; $4872$ with $200$ s PEs; $4806$ with $300$ PEs, $4724$ with $400$ PEs and $4611$ with $500$ PEs. Hence the administrators have to opt for either upgrading the hardware or running multiple instances of Controller if flow requests are to be generated at the rate of $5000$ requests/second. 
Although there is an overhead with the PbSA, note that the baseline mode will result in establishing the routes for all the flow requests (including malicious requests). However, with PbSA only secure routes will be established corresponding to security policy specifications. Hence for instance, during an outbreak of a worm attack, when the compromised hosts randomly scan for vulnerable machines for spreading the attack, the baseline mode will result in successful establishment of the routes for all the malicious flow requests. With PbSA all the randomly generated malicious flow requests which do not have corresponding PEs permitting the flows will be dropped. This is a significant advantage for overcoming attacks in networks, especially as the malicious flow requests are dropped at the source end. \\
\begin{figure}
\centering
\includegraphics[scale=0.33]{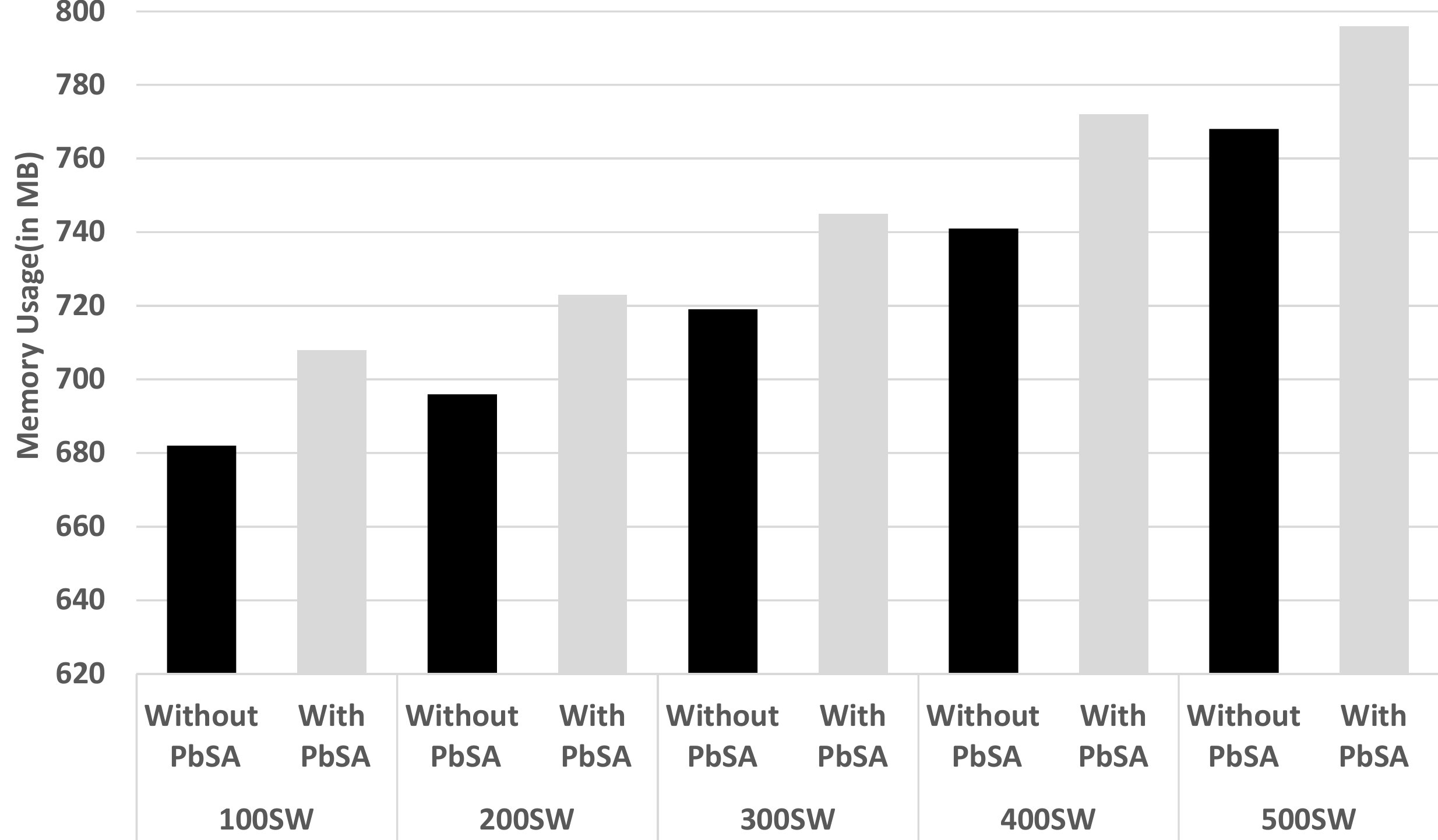}
\caption{Memory Overhead with PbSA }
\vspace*{-3mm}
\label{fig:mopbsa}
\end{figure}
\noindent We have also analysed memory overhead with PbSA with $500$ policy expresseions (PEs) with varying number of switches. 
As shown in Figure \ref{fig:mopbsa}, there is an increase in the memory usage with PbSA. Also the memory usage increases with the increase in the number of PEs stored in the PbSA.\\
\begin{figure}
\centering
\includegraphics[scale=0.30]{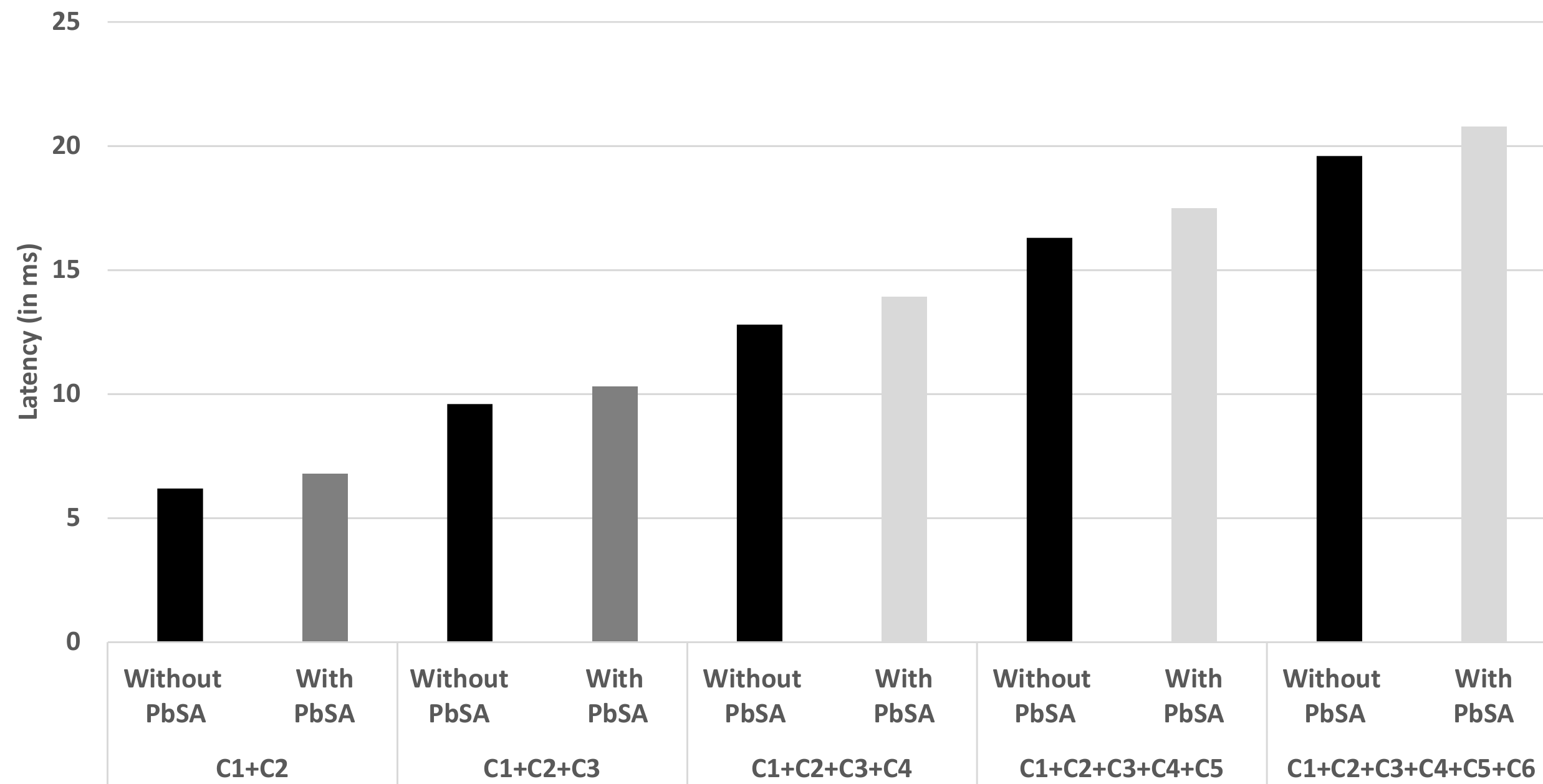}
\caption{Latency at the Controller for multi AS domains }
\label{fig:lcm}
\end{figure}
\noindent With multiple AS domains, we have considered the case of sequentially connected Controllers between the source and the destination hosts. When the source host sends an intent message to its Controller for communication with a destination host in another AS domain, the intent message is initially processed by the Controller in the source AS domain and then sequentially forwarded to the Controller of a directly connected AS domain. The process is repeated until the intent message is forwarded and processed by the Controller to which the destination host is connected. Similar to the intra domain situation, we have analysed the performance first in the baseline mode without the PbSA and then with PbSA enabled on all the Controllers. In the case of a multi AS domain scenario, there are three hops for each AS domain, namely ingress, transit and egress hops. The ingress hop receives a packet flow from the source end host or from an egress switch of the previous AS domain. There is one transit switch connecting the ingress switch to the egress switch. The egress switch forwards the flow request to the destination host or to an ingress switch of the next AS domain. We have analysed latency and end-to-end flow establishment in such a scenario.  \\
Figure \ref{fig:lcm} shows the latency results for a multidomain scenario. In such a scenario, there is latency in each Controller. The latency results shown are obtained by adding the latency at each Controller without the PbSA. Then we enabled PbSA on each Controller and obtained the results. Similar to the intra-domain scenario, there is an increase in the latency with the PbSA. Also, the latency increases with the increase in the number of AS domains in the baseline mode and with PbSA. However, note that with PbSA, the intent messages are forwarded to another Controller only if it is permitted according to the policies specified in the PbSA. All the malicious intent messages which do not have corresponding PEs to permit the intent messages are dropped at the first Controller itself. This leads to saving of resources in the Controllers in other AS domains.  
\begin{figure}
\centering
\includegraphics[scale=0.29]{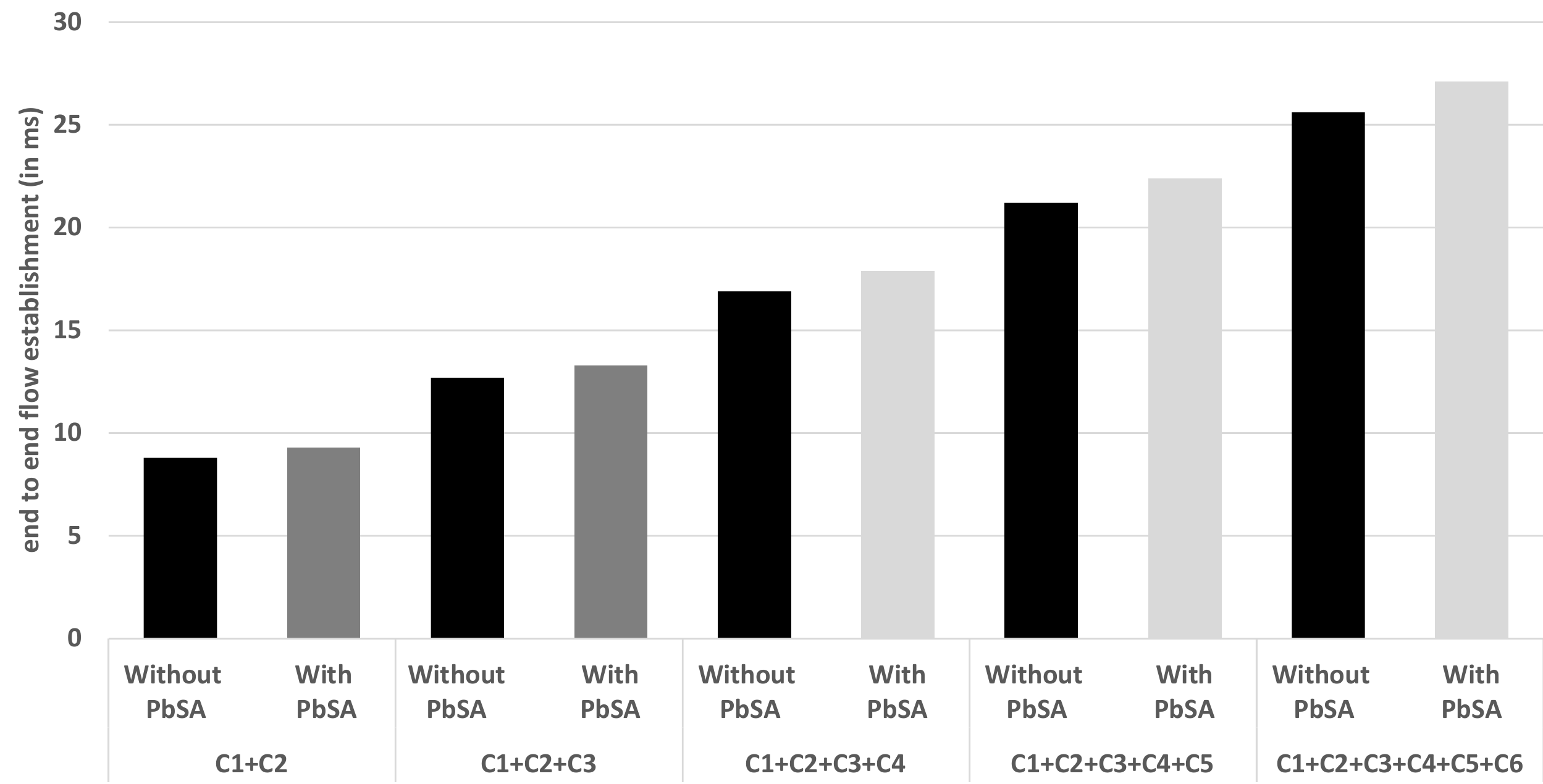}
\caption{End-to-End flow establishment time }
\label{fig:eefet}
\end{figure}
Figure \ref{fig:eefet} shows the end-to-end flow establishment time with and without the PbSA. Although there is some additional delay with the PbSA, note that such a delay is only applicable to valid flows. There is no need for flow establishment if there is no corresponding PE permitting the flow.  \\
\begin{figure}
\centering
\includegraphics[scale=0.30]{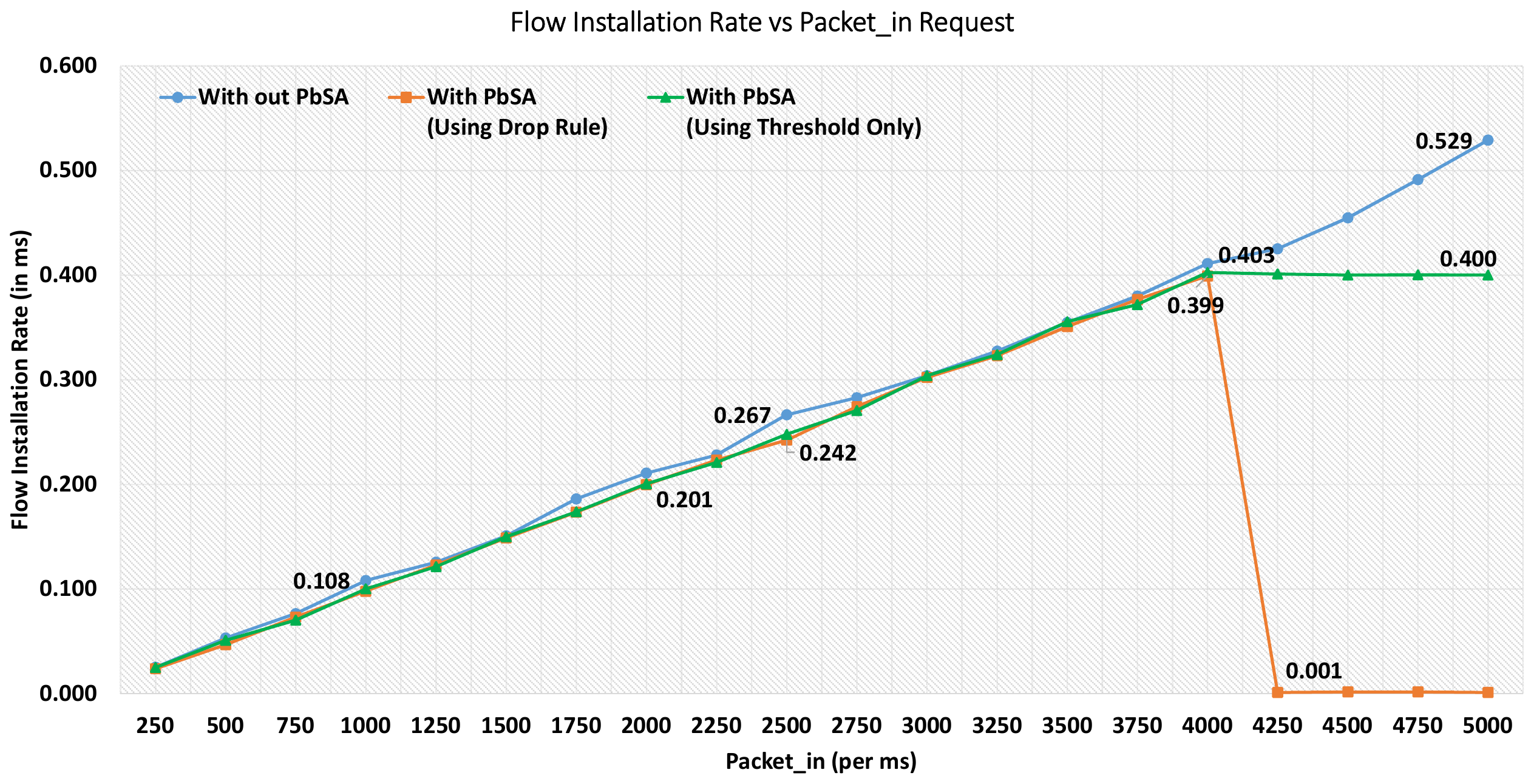}
\caption{Flow installation rate with varying Packet\_in Request }
\label{fig:fir}
\end{figure}
\noindent Now we consider the attack scenario (4) outlined in Section \ref{subsec:fld}. \\
First we used a single attacking host machine connected to an OpenFlow switch flooding the Controller with DDoS Packet\_in attack requests and measured the flow installation rate. 
Figure~\ref{fig:fir} presents 3 curves representing 3 different experimental scenarios: 1) Without PbSA active; 2) With PbSA active using a threshold policy and 3) With PbSA active using a drop rule policy. \\
When PbSA is not active, as expected, the number of flow installation rate increases with the increase in the number of Packet\_in requests (shown by the blue curve in the Figure~\ref{fig:fir}). For 250 Packet\_in request/ms, the flow installation rate is around 0.025295 flows/ms and at 5000 Packet\_in requests/ms the flow installation rate is around 0.52904. This incurs consumption of SDN Controller resources. \\
With PbSA active, with the threshold policy applied at 4000 requests/ms, the Controller limits the packet requests to 4000. The PbSA detects the flooding and above the threshold, it maintains a constant flow installation rate as per policy expression (to 4000 Packet\_in requests/ms). This is shown by the continuing green line in the Figure~\ref{fig:fir}. The Controller is not overloaded by malicious Packet\_in requests above 4000, and the flow installation rate remains constant around 0.40001- 0.402631 flows/ms.
In the third scenario, we enforced a drop rule policy above the threshold. In this case, the PbSA detects the flooding and it installs a drop rule (in the OpenFlow switch) specific to the malicious host. Hence all the subsequent packet requests from this malicious host are dropped. However there is a delay for this to occur and this corresponds to the time taken to install the drop rule at the switch and for this rule to be enforced. This is shown by the orange line in the Figure~\ref{fig:fir}. For 4000-5000 Packet\_in requests/ms, the flow installation rate drops to 0.0012 flows/ms and it remains constant around 0.0012 - 0.0016 flows/ms.
\section{Security Analysis}\label{sec:SA}
First in this section we examine how the security requirements that we originally set out in Section II are achieved by the proposed security architecture.\\ 
• (R1) The core aspect of our security architecture PbSA involves security policy based authorization of flows in distributed SDN environment. These security policies use different flow parameters to authorize flows in intra and inter domain communications in SDN. The policy administrators can define fine grained security policies based on different attributes such as users/devices/AS identities and services running on the end hosts, and context attributes such as time and location of devices as well as path based specifications bsaed on security labels of switches. \\
• (R2) PbSA is used for security management of devices in AS domains and enhancing the security of SDN operations. For instance, PbSA enforces a default deny policy and drops all the flow requests that do not satisfy policies permitting the flow. Threshold policies have been used to detect attacks such as worms and drop malicious flow requests when the infected machines generate flooding attacks to destinations. \\   
• (R3) PbSA makes use of the network domain wide information available at the SDN Controller and the security policies stored in the policy repository for achieving end to end security within a domain. PbSA is based on modular design to enable incremental implementation of the required security modules, making it easily extendable with additional security features. For instance, we have implemented modules for path selection based on the security labels of the switches, and attack detection using signature and thresholds, and are currently developing key management modules for generating and distributing keys for securing the communication between network devices.\\ 
• (R4) Our PbSA architecture can be used to secure end to end communication across multiple AS domains.  A novel feature of the proposed security architecture involves the use  of  the  dynamic  visibility  of  the  network  connectivity  to specify flow and path based security policies to achieve secure communications  and  efficient  provision  of  services  across multiple  SDN domains. The specific flow requirements are validated against the policies of all the AS domains (source, transit and destination) before the routes are established to enable authorized communication between the end hosts. For instance, suppose due to a DDoS attack, traffic from an  end  host is  not  able  to  get  through  the  network.  PbSA is able to detect the DoS attack efficiently  and establish an alternative  path  for  the  traffic from the end host to the required destination.  Moreover,  we  have shown that PbSA can enforce  policies  such  as  certain  communications  should  go  through  a  path  with  certain security attributes. Such path based policies are critical when securing data from sensitive applications but are also useful for applications with different quality of service requirements. \\

\vspace{-5mm}

Let us now consider how the proposed security architecture deals with the attacks mentioned earlier in Section II.\\  
•	(A1) Our security architecture enforces default deny communication between the hosts. If the end hosts are located within the same domain, then the communication is permitted if it meets the policy requirements of the AS domain. In this case, the unauthorised communication requests are dropped at the SDN Controller within the domain. If the end hosts are located in different domains, then the communication is permitted if it meets the policy requirements of all the AS domains involved namely the source AS, transit AS and destination AS domains. For instance, the scan messages that are destined to random destination addresses during the spread of worms will not result in establishment of routes to the destination host. In this case the unauthorised communication requests will be dropped at the AS domains where authroization policies are not satisfied.\\
•	(A2) The policy repository in the PbSA has information about the services hosted on the servers in the domain. The security policy administrators can specify the usage policies based on different parameters such as the users/devices/AS identities and the services running on the end hosts, time of flows and location of devices. When an end host attempts to access an unauthorised service, the switch that is connected to the malicious end host will generate a packet_in message to the SDN Controller. Since the flow request will not be authorised by the PbSA, the malicious flow request will be denied.\\
•	(A3) As security policy administrators can specify policies based on different parameters such as location of the devices and AS domains, attacks originating from hosts residing in certain locations are prevented by PbSA.  \\
•	(A4) If an end host is directly targeting any of the switches, then we have shown that PbSA is able to detect such malicious attacks. For example, a malicious end host attempting to flood a particular switch in the network will be prevented by policies in PbSA detecting such attacks. Also, the attackers do not have any control on the path taken by the malicious traffic since the routes are established dynamically for each flow request based on policy specifications. If there is a specific need for the end host traffic to take a specific path, then this has to be conveyed in the intent message and satisfy the policies in the PbSA. \\
•	(A5) In our current design, we have used domain constraints to specify threshold policies over multiple domains. This has been used to enforce restrictions on new flows. Consider an attack that chains malicious flows coming from different locations.  Recall that each Controller has a global view of all the established routes and active flows within its domain. Also, each flow request is validated by the Controllers in all the relevant domains before the flow is permitted. If the traffic at any link is above the threshold specified in the policies, then the Controller of that domain can select an alternative path to the destination. If there is no alternative path to the destination, then the Controller will not accept this new flow request. Furthermore, if any of the end hosts within a domain is generating a flow request above the threshold, then the switch will drop the flows from the maliciuos host and the Controller will dynamically configure rule in that switch. Hence, the malicious flow request will not result in the establishment of the new routes, and the attack that attempts to chain malicious flows from different locations are prevented. However the use of such domain constraints across multiple domains is restrictive; in our future implementation, we will be relaxing this restriction using a hierarchy of Controllers and using policy inheritance to better coordinate the policies in Controllers in different domains.

\vspace{-0.01in}

\section{Related Work}\label{sec:RW}
Inter-domain routing forms the main backbone in today's network infrastructures such as the Internet. There are some 50,000 autonomous domains. A breakthrough in recent times in the networking world is the paradigm of SDN and the programmability, flexibility, and manageability that SDNs offer. This paper has proposed a security policy based control for both intra and inter-domain communications in a distributed SDN environment, which offers significant advantages both in terms of specification of authorized flows as well as detection and prevention of attacks in networks. In this section, we will discuss different related works that are relevant to our research. 

\vspace{0.03in}

Kreutz et al. in ~\cite{kreutz13} presented the seven threat vector in SDN space. Our threat model has taken this as the basis and enhanced it in the context of intra and inter-domain communications in SDNs. In particular, we have shown how a policy based security architecture can be used to counteract flow related specific threats discussed in ~\cite{kreutz13}. Furthermore, ~\cite{kreutz13} advocates the need to consider security and dependability in such networks. Currently we are in the process of developing a trust model for SDN, which addresses trust relationships between devices, switches, controllers and their applications; the proposed policy based security architecture together with the trust model will help to achieve security and dependability properties in SDNs.

\vspace{0.03in}

Clark in RFC1102 described policy-based routing for legacy inter-domain networks~\cite{clark89}. He proposed a language based approach to policy specification to control routing of packets within autonomous system domains. This work formed the basis of our security policy language in our security architecture. Tsudik \textit{et al.} in~\cite{tsudik89} introduced security measures to policy-based inter-domain routing. Their work was mainly concerned with key management using symmetric keys to secure communications in AS domains. We have used Clark's policy language syntax and extended it to develop fine grained security policy specifications in SDN for both intra and inter domain communications. We have developed a security architecture which enables the enforcement of these security policies in the Open Flow switches on the traffic originating from the end hosts and users connected to these switches. As mentioned before, we will be extending this security policy based architecture with key management to secure communcations using a similar approach to that suggested in ~\cite{tsudik89}. As indicated earlier, this extension on key management is traditional and well-known. The novelty of our proposed security architecture is the use of policy based specifications enabling fine grained path and flow based policies for multiple domains as well as the ability to capture dynamic aspects of the network and flow context within these policy based specifications enabling the detection of attacks. This extends our previous work on policy based approach for intra-domain communications ~\cite{karmakar2016} and develops a comprehensive security architecture for end to end services across multiple domains.

\vspace{0.03in}

Peter \textit{et al.} in ~\cite{peter13} discussed about BGP issues in SDN and proposed an application-based routing architecture. OpenDaylight ~\cite{odl} and ONOS ~\cite{onos} have followed this approach but have limited features to establish communications between SDN Controllers in autonomous systems using BGP. OpenDaylight uses SDNi protocol ~\cite{sdni} to exchange state information between domain Controllers. On the other hand, ONOS uses SDN-IP ~\cite{sdnip} to communicate with other autonomous system SDN Controllers. The core features of the BGP  such as no explicit support for iBGP sessions, IPv6 and limitations on the number of routes ~\cite{sdnip} are the major weaknesses of SDN-IP. 
Routing Control Platform (RCP) ~\cite{rothenberg12}  for SDNs uses BGP route speakers which are different from IP forwarding plane. Our PbSA for SDNs also uses application-based approach and emphasizes the need for policy-based routing for both intra and inter-domain communications in SDNs.

\vspace{0.03in}

Fresco is a modular security framework designed explicitly for NOX controllers \cite{shin13fresco}. It provides a development environment which allows modules to be developed in terms of scripts that can help to detect specific attacks such as threshold based attacks, and drop or quarantine the malicious packets. Their focus is mainly on the framework and development environment, particularly aimed at a single domain. Our approach in contrast considers a security architecture that allows security policy specifications for controlling flows in both intra and inter domain SDN environments as well as enabling authorized access to services across multiple domains. Furthermore, our architecture enables fine grained flow and path based policies that are relevant for critical applications. As part of our architecture implementation, we will be exploring the use of Fresco Development Environment (FRESCO DE) to enable automatic translation of our flow policies into rules in the OpenFlow enabled switches.

\vspace{0.03in}

SDN Controllers such as Maple\cite{ voellmy13}, Nettle \cite{ voellmy11} and flow languages like Flow-based Management Language (FML)~\cite{hinrichs05},  Frenetic~\cite{foster11} and Pyretic~\cite{reich13}  provide ways to express high-level network policies using Haskell and Python to manage intra-domain SDN interactions. Lee \textit{et al.}~\cite{lee2017security} uses SDN based policy management functions to logically separate the network services thereby decreasing the attack surface and then combining them to create composite services. Our approach is different from the one proposed in ~\cite{lee2017security}; however, in principle, we can also take advantage of such an approach and partition our security policy specifications based on, for instance, virtualised services.  
Sahay \textit{et al.} in ~\cite{sahay2017adaptive} proposed a dynamic policy enforcement framework that allows ISPs to specify security policies to mitigate the impact of network attacks by taking into account the specific requirements of their customers. Their focus is on the use of SDN to gather information from ISP network devics to detect and mitigate attacks. Again this is different to our architecture which is concerned with policy based specification of authorized flows in intra and inter domain SDN as well as use of these policies in the detection and mitigation of attacks.

\vspace{0.03in}
Some other related works include the following: NetPlumber, which is concerned with checking the reachability properties of flows in OpenFlow Network \cite{kazemian13rtnp}; Athena, which is a distributed anomaly detection framework for detecting packet anomaly in SDN ~\cite{lee17athena}; and the work in ~\cite{Deng17inject} considering the use of OpenFlow Packet\_in messages to generate topology poisoning attacks. Though they involve security attacks and SDN, they are not directly related to our work on security policy based management and provision of secure end to end flows and services across multiple domains in SDNs. 
\vspace*{-4mm}

\section{Conclusion}\label{sec:C}

\vspace*{-1mm}

\noindent In this paper, we have presented a policy based security architecture for distributed SDN, enabling secure intra and inter-domain communications and flows between different end hosts across multiple domains. 
Our security architecture uses a policy language based approach to specifying security requirements at a fine granular level to control the flow of information in a multi-domain SDN environment. In large networks with multiple autonomous domains crossing organizational boundaries, such a policy based security architecture provides a convenient way for dealing with data flows across different jurisdictions satisfying different constraints. In particular such an approach is useful when it comes to flow of sensitive information and traffic which need to satisfy specific routing and path security constraints, in addition to traditional congestion and cost requirements. Our architecture is able to specify various security policies at fine granular level, based on a variety of attributes of users, devices/ switches, services, location as well as security labels associated with the switches and Controllers in different domains. An important characteristic of our architecture is its ability to specify path and flow based security policies, which is a distinct advantage for SDN based services and applications. This enables us to specify requirements such as certain flows need to go through a path of switches satisfying certain constraints based on different attributes.  This also enables to define prohibited paths for certain types of traffic or even the need for certain switches to be obligatorily traversed for certain flows. Such path based policies are not only relevant in security critical applications but also useful in normal applications which may have different requirements for different types of traffic. We have analysed the performance characteristics of our architecture as well as discussed how the architecture is able to counteract various security attacks and meet the different security requirements using the policy based mechanisms. The ability to distribute security capabilities intelligently as a service layer and to have a dynamic security policy based approach to securing a multitude of devices against attacks are important contributions of this paper.

\vspace{0.03in}

As further work, we are exploring several extensions to our architecture. First, we are currently in the process of extending the implementation of intra-domain key management to inter-domain key management. To achieve secure communications, our earlier architecture  ~\cite{karmakar2016} supports on demand confidentiality and integrity of traffic at the switches. The key management module that we had developed in the Controller employs symmetric keys for securing the communications within a domain. We had extended the OpenFlow protocol to distribute the keys in a secure manner. We had also extended the functionality in the switches to receive the keys. We have not described this key management functionality of our architecture in this paper due to space considerations. We are currently extending these key management mechanisms between different SDN Controllers to achieve secure communications in an inter-domain setting. This involves the traditional hybrid key management scheme with both public key and symmetric key based techniques. This is being done to achieve completeness in the security architecture. Second, we are developing a formal model that captures the essential features of SDN behaviour and are introducing a formal language for specifying intra and inter domain flows considered in this paper. We plan to specify the security policies  proposed in this paper formally for this SDN model and develop proof techniques for validating the behaviour of the SDN model against the security policies. Using this method, we plan to formally prove certain security properties derived from the policies, which we believe will be helpful to demonstrate a higher level of assurance of the policy based security architecture enabled SDN applications and services. 

\vspace{0.05in}
%
\begin{small}
\bibliographystyle{IEEEtran}
\bibliography{IEEEabrv,biblcn}
\end{small}
\begin{IEEEbiography}[{\includegraphics[width=1in,height=1in,clip,keepaspectratio]{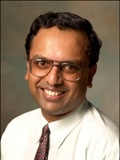}}]
{Vijay Varadharajan}
is the Global Innovation Chair Professor in Cyber Security at the University of Newcastle. He is also the Director of the Advanced Cyber Security Research Centre. Vijay has published more than 400 papers in international journals and conferences. Vijay has been/is on the Editorial Board of several journals including ACM TISSEC, IEEE TDSC, TIFS and TCC.
\end{IEEEbiography}
\vspace{-50pt}
\begin{IEEEbiography}[{\includegraphics[width=1in,height=1in,clip,keepaspectratio]{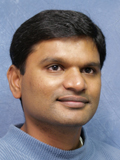}}]{Uday Tupakula}
received the PhD degree under the supervision of Prof. Varadharajan, in 2006. He is a senior lecturer with the University of Newcastle, Australia. Uday has 75 publications in different research areas such as network security,DDoS attacks, MANET security, and secure virtual systems. He is a Member of the IEEE, BCS and ACM.
\end{IEEEbiography}
\vspace{-50pt}
\begin{IEEEbiography}[{\includegraphics[width=1in,height=1in,clip,keepaspectratio]{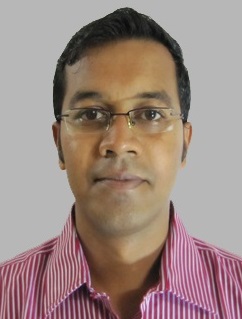}}]{Kallol Karmakar}
is a Ph.D. student at the Advanced Cyber Security Research Centre. He is working on Software Defined Network security. 
\end{IEEEbiography}
\vspace{-50pt}
\begin{IEEEbiography}
[{\includegraphics[width=1in,height=1in,clip,keepaspectratio]{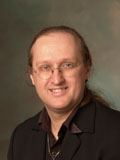}}]{Michael Hitchens}
is the Associate Dean of Quality and Standards in the Faculty of Science, Macquarie University. His research interests access control, security protocols, trust in distributed systems and game design. 
\end{IEEEbiography}
\end{document}